\documentclass[runningheads,a4paper]{llncs}
\usepackage{amssymb}
\usepackage{graphicx}
\usepackage[]{amsmath}
 
\usepackage{url}
\urldef{\mailsa}\path|{gba, sangelini, mbianchi,lcostantini, gmarcone}@fub.it| 
\newcommand{\keywords}[1]{\par\addvspace\baselineskip
\noindent\keywordname\enspace\ignorespaces#1}

\usepackage[english]{babel}
\usepackage{lscape}
\usepackage{subfigure}
\usepackage[boxed]{algorithm2e}
\usepackage{color}
\definecolor{darkgreen1}{rgb}{0,0.6,0}
\definecolor{darkgreen}{rgb}{0,0.4,0}
\definecolor{lightpink}{rgb}{1,0.9,0.9}
\definecolor{lightblue}{rgb}{0.8,1,1}
\definecolor{darkblue}{rgb}{0,0,0.8}
\definecolor{darkmagenta}{rgb}{0.5,0,0.5}
\definecolor{darkred}{rgb}{0.6,0,0}
\definecolor{giallino}{rgb}{0.5,0.5,0}
\definecolor{flashgreen}{rgb}{0,0.9,0.9}
\definecolor{violetto}{rgb}{0.8,0,0.8}

\newcommand {\term} {t}
\newcommand {\doc} {d}
\newcommand {\category} {c}

\newcommand {\tf} {\textit{x}}

\renewcommand {\P} {\textrm {P}}

  \newcommand{\stress}[1]{{{\bf #1}}}

\newcommand {\query} {q}
 
\newcommand {\Elitet} {\stress {E_ \term}}
\newcommand {\length} {  l  }

\newcommand {\FreqTotCollElite} 
{\stress{\sum_{d'\in\Elitet} length(d')}}

 \usepackage{array}
\newcolumntype{L}[1]{>{\raggedright\let\newline\\\arraybackslash\hspace{0pt}}m{#1}}
\newcolumntype{C}[1]{>{\centering\let\newline\\\arraybackslash\hspace{0pt}}m{#1}}
\newcolumntype{R}[1]{>{\raggedleft\let\newline\\\arraybackslash\hspace{0pt}}m{#1}}

\begin{document}

\mainmatter

\title{A cumulative approach to quantification for sentiment analysis}

\titlerunning{A cumulative approach to sentiment analysis}

\author{Giambattista Amati
\and Simone Angelini\and Marco Bianchi\and\\Luca Costantini\and Giuseppe Marcone
}
\authorrunning{A cumulative approach to sentiment analysis}

\institute{ Fondazione Ugo Bordoni, Viale del Policlinico 147,  00161 Roma}

\toctitle{A cumulative approach to sentiment analysis}

\maketitle

\begin{abstract}
\emph{ We estimate sentiment categories proportions for retrieval within large retrieval sets. In general, estimates are produced by counting the classification outcomes and then by adjusting such category sizes taking into account misclassification error matrix. However, both the accuracy of the classifier and the precision of the retrieval produce a large number of errors that makes difficult the application of an aggregative approach to sentiment analysis as a reliable and efficient estimation of proportions for sentiment categories. 
 The challenge for real time analytics during retrieval is thus to overcome misclassification errors, and more importantly, to apply sentiment classification or any other similar post-processing analytics at retrieval time. We present a non-aggregative approach that can be applied to very large retrieval sets of queries.
}
\end{abstract}

\section{Introduction}
   We study the problem of  estimating the size and  proportions  of  sentiment categories   (category quantification  \cite{ecml/Forman05,datamine/Forman08,BarranqueroDC15,BellaFHR14}) over a result set of a   query.  The   quantification problem is very challenging because of several factors: the sentiment content drift caused by the content of a query, the size of the result set, the term sparsity, the precision of the retrieval,  finally the accuracy of the classifier. In sentiment quantification the number of classification errors (false positives and false negatives) as well as for each result set  the ability of the classifier to balance the priors of the sentiment categories   are both important. Indeed, existent test sets show that both error rates and categories priors may largely vary each  topic or result set.
There are four possible approaches to quantification:  to adjust   counts with a confusion matrix,  to choose a suitable training set to learn the classifier in order to   better fit the priors to the new data, to improve the quantification accuracy with a proper multivariate classification model, to smooth  the classification counts with a second learning model.  

The first approach  (the AC\&C approach) classifies documents in the retrieval set  $D_\query$  over a certain number of categories $\category$, and then counting of the elements (the set $\hat c$) falling into each category $\category$ is eventually adjusted to the final estimate $\hat c\cap D_\query$ with the numbers of the misclassification errors that the classifier makes on a training set and that is provided by the confusion matrix $p(\hat c_i|c_j)_{i\neq j}$ \cite{ecml/Forman05,datamine/Forman08}. Among these approaches the empirical median sweep find  exhaustively any  possible classification threshold to obtain an estimate of the prevalence. The final prevalence quantity is the median of all the estimates.

The second approach is to use a set of spanning features, that must be drawn randomly and  independently from  the categories, that is then used to  draw a suitable training   sample from a validation set. Not all the validation set is used to train the classifier but a proper subset.  The drawn training set  turnes out to be the closest set to the collection according to a distance, for example  the Kullback-Leibler Divergence or the Hellinger distance \cite{GonzaLez-Castro:2013}. Such  distance  is between   the   two distributions of the features: the first  on the collection,  the second on the retrieval set.  Though the  Hopkins and King method is not automatic \cite{HopKin10}, it can be still fall into such an approach, since its smooths the raw estimates of a manual evaluation by counting categories over a spanning set of features in the collection.

Since quantification accuracy is related to the ability of the quantification model  to minimize the difference in size of false positives and false negatives,  at certain extent the classification accuracy is independent from the quantification accuracy. However, it is also true that the higher the classifier accuracy is, the less    the  difference  in size of the errors is,  all other experiment settings remaining the same. One obstacle to achieve a higher quantification accuracy   is that some classifiers are binary in nature (such as SVM or  the approach based on the Hellinger Distance) so a different  approach has shown to achieve a better quantification accuracy  under a multivariate approach \cite{joachims05multivariateperformance}. With a higher  accuracy the multivariate approach   avoids smoothing methods based on the confusion matrix.

The last approach is non aggregative and use two distinct learning models: the first is the classifier the second model learns how to quantify from the classifier. Instead of using the confusion matrix this approach does not use the classifier as a Bayesian decision rule but cumulates the scores used to emit such  decisions and correlates observed categories sizes to such aggregate scores through for example regression models \cite{DART2014}.      

For particular dataset, such as  the Internet Movie  Dataset, there is also a link-based quantification model \cite{Diakopoulos:2010:CDP}, and an iterative method \cite{Xue:2009:QSC} to rectify  the classifier when the prevalence of a class may change over time. The Expectation Maximization can be also applied for adjusting the outputs of a classifier with new  class priors\cite{Latinne01adjustingthe}.

We introduce a  non aggregative approach on Sections \ref{sec::cumulative::classifier} and \ref{sec::cumulative}. We define the experimental settings and the evaluation measures suitable in a retrieval scenario. In particular we use the Kolmogorov-Smirnov distance, and its p-value also provides a statistical significance test for validating the goodness-of-fitting of the new model.

\section{Related Works}
According to the family of the Adjusted Classify \& Count methods, once the   classifier  returns a set $\hat \category$ for each category $c$ in a proportion $P(\hat \category_j|\query)$ among the $n$ categories, the {\em Theorem of Total Probability}   decomposes these classifier outcomes over the set of $n$  categories \cite{LevyKass:1970} [$\P(\hat \category_j|\query)=\sum_{i=1}^n{\P(\hat c_j|\category_i,\query)}{\P(\category_i|\query)} \textit{ j=1,\ldots, n}$].
 The Scaled Probability Average  approach is a variant of the ACC method, with the  expectation over the categories probabilities used instead of the   total probability theorem\cite{BellaFHR14}.  
 The unknown estimates $P(\category_i|\query)$ are  derived    solving a simple linear system of $n$ equations with $n$ variables:  
${{\underset{n\times n}{\P(\hat\category|\category,\query)}}\cdot {\underset{n\times 1}{\P(c|\query)}}=\underset{n\times 1}{\P(\hat\category|\query)}}$. 
 The accuracy of the classifier should not  matter, since the misclassification errors are used to  estimate all category sizes.
 This model can be easily extended with a linear regression model to learn from a set of queries (or different training sets), i.e.
${{\underset{n\times n}{\P(\hat\category|\category,\query)}}\cdot {\underset{n\times Q}{\P(c|\query)}}\sim\underset{n\times Q}{\P(\hat\category|\query)}}$ 
  or with an entropy value $H$ substituted for $\P$\cite{DART2014}, ${{\underset{n\times n}{H(\hat\category|\category,\query)}}\cdot {\underset{n\times Q}{\P(c|\query)}}\sim\underset{n\times Q}{\P(\hat\category|\query)}}$. 
Here, $\sim$ stands for equality up to linear regression coefficients that  fit the the model with $|Q|$ equations.

\section{Cumulative Classifiers \label{sec::cumulative::classifier}}

\label{sec::classifier}
We use the learning models of three classifiers,  MNB, SVM and DBM, and apply  a cumulative measure \begin{equation}
\mu_\category(\sum_\doc\tf_i| \tf_i \textrm{ frequency of } i \textrm{ in } \doc, \doc\in D_\query)\end{equation} for the retrieval set $D_\query $ and category $\category$ of documents, that is a measure satisfying the following property:
 $\mu_\category (\sum_i X_i) =\sum_{\doc \in D,i}\mu_\category(\tf_i)$ with $X_i=\sum_{\doc \in D}\tf_i$ and $\sum_i\mu_\category(\tf_i)$ used to classify documents $\vec x$. Such an additive property derived from a classifier, allows us to make the hypothesis  that   the cumulative function $\mu_\category(D_\query)$ correlates (linearly) with the number of documents that   are  relevant to the query ($\vec x\in R_\query$) and are   in the category $\category$:\begin{equation}
 \Phi_\category(\{\mu_\category\}_{\category\in{\cal C}}, D_\query, \theta) = |R_\query\cap\category|\label{Eq:cumulative}
 \end{equation} 
 Obviously not all classifiers are suitable for defining  such a cumulative measure, but MNB, DBM and SVM are.

Since the learning probabilistic model of MNB  is based on the term independence assumption,  the logarithm of probabilities is additive over terms. The Kullback-Leibler divergence  is also additive \cite{Kullback59}, so that both the probabilistic learning model of DBM  and   MNB satisfy the additivity  property over independent terms. Analogously,    
SVM can be seen as  cumulative measure function with respect to the direction of the hyperplanes because distance is additive along that direction. We now show that these additive properties  are necessary  conditions in order to derive a cumulative measure for these three classifiers. Moreover,
Table \ref{table::correlation} shows that there  is a linear correlation between  the cumulative  measures   of MNB and DBM over the categories and the cardinalities  of their respective categories (positive versus negative). 
\\{\bf Multinomial Naive Bayes (MNB)}. 
Due to the sparsity of data, Naive Bayes (NB) and gaussian Naive Bayes perform poorly in text classification\cite{Rennie2003}, therefore the MNB classifier is preferred to NB.
Let $\tf_i$ be the frequency of word $i$ in document $\doc$, and $p(\category)$ be the prior for  category $\category$, that is the frequency of  category $\category$ in the training  collection, and $f_{i,\category}$ the frequency of the word $i$ in category $\category$ containing $L_\category$ tokens of words.  Most of the implementations of MNB \cite{Manning:2008:IIR}
 maximize the  logarithm of the likelihood with a multinomial distribution (one for each category) as follows:
\begin{eqnarray}
\arg\max_{\category}\left[ 
\log p(\category) + \sum_{i }\tf_i\log\left(\frac{f_{i,\category}+\alpha_i}{L_\category+\alpha}\right)\right]\label{Eq:MNB:Dec}
\end{eqnarray}
where $\alpha=\sum_i \alpha_i$ and $\alpha_i$ smoothing parameters. We choose $\alpha_i =1$ \cite{Rennie2003}. The cumulative function is \begin{eqnarray}
\mu_\category(\vec x) = \sum_{i } \tf_i\log\left(\frac{f_{i,\category}+\alpha_i}{L_\category+\alpha}\right)\label{eq:cumulative:MNB}\end{eqnarray}
It is easy to verify that $ \mu_\category(D) =\sum_{\vec x \in D}\mu_\category(\vec x) $.
\\ {\bf  SVM classifier}.
 SVM constructs a direction
 $\vec w=\sum_j \alpha^j y^j {\vec x}^j
\label{eq::svm}$, 
$\vec x^j$ being the vector containing all the frequencies of the $j$-th support document.
 The distances of documents $ \vec x$, considered as vectors of terms,   from  the hyperplanes of equations $\vec w\cdot \vec x + b = 1$ and $\vec w\cdot \vec x + b =-1$ define a decision rule to assign a document to a category.  Differently from  probabilistic learning models, where we can exploit the additivity property of the logarithm function over independent events, we here make use of  the additivity property of distance along the  normal direction $\vec w $ to both category hyperplanes. Then, we   assume that the sum of the distances of documents  from an hyperplane, that is $\sum_{\doc\in D} \mu_\category(\vec x)$,  is linearly correlated to the number $|\category|$ of the elements in the corresponding  category, that is such an assumption is $\sum_{\doc\in D} \mu_\category(\vec x)\propto|\category|$. Therefore, if $\mu_\category(\vec x) =\vec w \cdot \vec x+b$, then
 the sum $\sum_{\doc\in D} \mu_\category(\vec x) = \sum_{\doc\in D} (\vec w \cdot \vec x+b) $ with  the constraint $(\vec w \cdot \vec x+b) >1\; (\textrm{or }\vec w \cdot \vec x +b  <-1 \textrm{ respectively}) $  provides a number $|\hat\category|$ of  positive (negative) documents in the result set $D$. 
 We note that  $\mu_\category(D) = \sum_{\doc\in D} \mu_\category(\vec x)$ up to the additive constant   $b\cdot |\hat\category|$. The distributive property of inner product  with respect to the sum of vectors $\vec x$ implies the cumulative property of $\mu_\category$  up to an additive constant proportional to $|\hat\category|$. However, the hypothesis of correlation between the cumulative function  $\mu_\category(D)$ and the estimated number $|\hat\category|$ of elements of the category is not affected because  $\mu_\category(D)\propto |\hat\category|$ is equivalent to $\mu_\category(D)+b\cdot |\hat\category|\propto |\hat\category|$. 
\\ {\bf Divergence-Based Model, DBM}.
The divergence based model DBM is a variant of the MNB of Equation \ref{Eq:MNB}. 
The difference with MNB  (Equation \ref{Eq:MNB}) lies in considering the factorials part in the calculation of the likelihood probabilities  \cite{Renyi69}.
 The multinomial distribution
with  a prior probability distribution $\pi_i$  over $L$ tokens can be approximated using the Kullback Leibler divergence.  If the frequency of the term $i$ in a category $\category$ equals  $f_{i,c}=\left\{ \frac{\sum_{\doc\in \category}x_{i,\category}}{L}\right\}$ when  $x_{i,\category}$ is the frequency of the term $i$ in document $\doc\in\category$, then an approximation of the multinomial is:
\begin{eqnarray}
 \sum_{i } f_{i,c}\cdot\log\left(\frac{f_{i,\category}}{\pi_i}\right) 
\label{Eq:MNB}
\end{eqnarray}
leading to the decision rule for a document $\vec x$:
\begin{eqnarray}
\arg\max_{\category}
\left[ 
\log p(\category) +\sum_{i }\tf_i\cdot f_{i,c}\cdot\log\left(\frac{f_{i,\category}}{\pi_i}\right)\right]
\label{Eq:MNB:classifier}
\end{eqnarray}
Note that, if $\length=\sum_i x_i$, $\sum_i f_{i,\category}\cdot\log\left(\frac{f_{i,\category}}{\pi_i}\right)= \length\cdot D(f_\category|\pi)$ with $D(f_\category|\pi)$ the Kullback-Leibler divergence  between the distributions $\{f_{i,\category}\}$ and $\{\pi_i\}$. Each token of the term $i$ contributes with  Formula (\ref{Eq:MNB}) in the decision rule. The model is learned on a training sample of $L$ tokens,   drawing the frequencies $f_{i,\category}$. The cumulative function is \begin{eqnarray}
\mu_\category(\vec x) = \sum_{i }\tf_i\cdot f_{i,c}\cdot\log\left(\frac{f_{i,\category}}{\pi_i}\right)\label{eq:cumulative:DBM}\end{eqnarray}
It is easy to verify that $ \mu_\category(D) =\sum_{\vec x \in D}\mu_\category(\vec x) $.

\section{Cumulative Quantification Models}
\label{sec::cumulative}
The cumulative approach requires a learning model $\Phi$ to correlate the cumulative function of a classifier $\mu_\category$ with the category size $\category$ as shown on Formula \ref{Eq:cumulative}.
Essential  parameters of $\theta$  are the size of $D_\query$,   the sparsity of the terms, that is correlated to the size of the lexicon, that in turns growths  following   Heap's Law  as long as new documents are indexed\cite{LeijenhorstVanderWeide05,Mandelbrot61}. The size of the lexicon  has a direct effect on the accuracy of the classifier. In a practical retrieval scenario one should expect to search, classify and count even millions of documents in one single shot so we need to verify whether quantification approaches can  scale both in effectiveness and efficiency.

We first observe that there exists a strong {\em linear}  correlation factor between cumulative sentiment $\mu_\category$ and category size. This hypothesis is statistically significant, as shown on  Table \ref{table::correlation}. In force of this evidence, we may initially set $\Phi$ to  a linear regression model. 
However, we encounter the following problems:\\
a)  Cumulative DBM and MNB learning models provide estimates of each category size that are learned independently from each other. Cumulative SVM instead uses both the two mutually exclusive categories in the training set, and   the quantification methods based on the Hellinger distance \cite{GonzaLez-Castro:2013} are only  suitable for binary classification.  In actual situations  there are at least six categories (non-relevant, positive, negative, mixed, neutral and other). Since they are mutually exclusive they should satisfy the constraint that $\sum_\category\mu_\category(D) = |D|$. Some of these categories are too difficult to be learned (such as the category of non-relevant document,  the neutral or the ``other'' category). Moreover many documents contain both opposite sentiment polarities, therefore the mixed category $M$ is  mutually exclusive but difficult to separate in practise from positive and negatives ones, $P$ and $N$.  To address multiple categories classification, the multivariate  SVM can be used instead of the confusion matrix. The training set is passed  as one single input to the multivariate  SVM and processed in one single  pass. According to \cite{GaoSebastiani:ASONAM2015}, the accuracy of multivariate SVM with a counting and classify approach shows  a better performance than the adjusted  counting and classify with confusion matrix and other variants.\\
b)  The learning model for $\Phi$ can be trained either by pooling all examples  irrespective of the set of queries used for the retrieval evaluation (item-driven learning), or can be trained  by aggregating  results query-by-query  (query-driven learning). In practise we train $\Phi$ with  an equation either for each observation or for each query.\\
c)    In order to normalize categories  one can include the size of the result set $D_\query$ as a parameter of a regression model. However, the regression model has several outliers, that occur  when their result set is very large $D_\query$. 
d)  In real applications  there exists a  high variability of the categories priors.  Besides the variability of the priors,  the size of test data is very small in all available data sets.  In conclusion, there always exists a mismatch between the training data distributions and the actual data distributions to which apply the quantification learning model.

\begin{figure}[!htpb]
    \centering
 \begin{subfigure} {\centering a) \label{fig::DIC::P::qd::PER}}
   \includegraphics[width=0.25\textwidth]{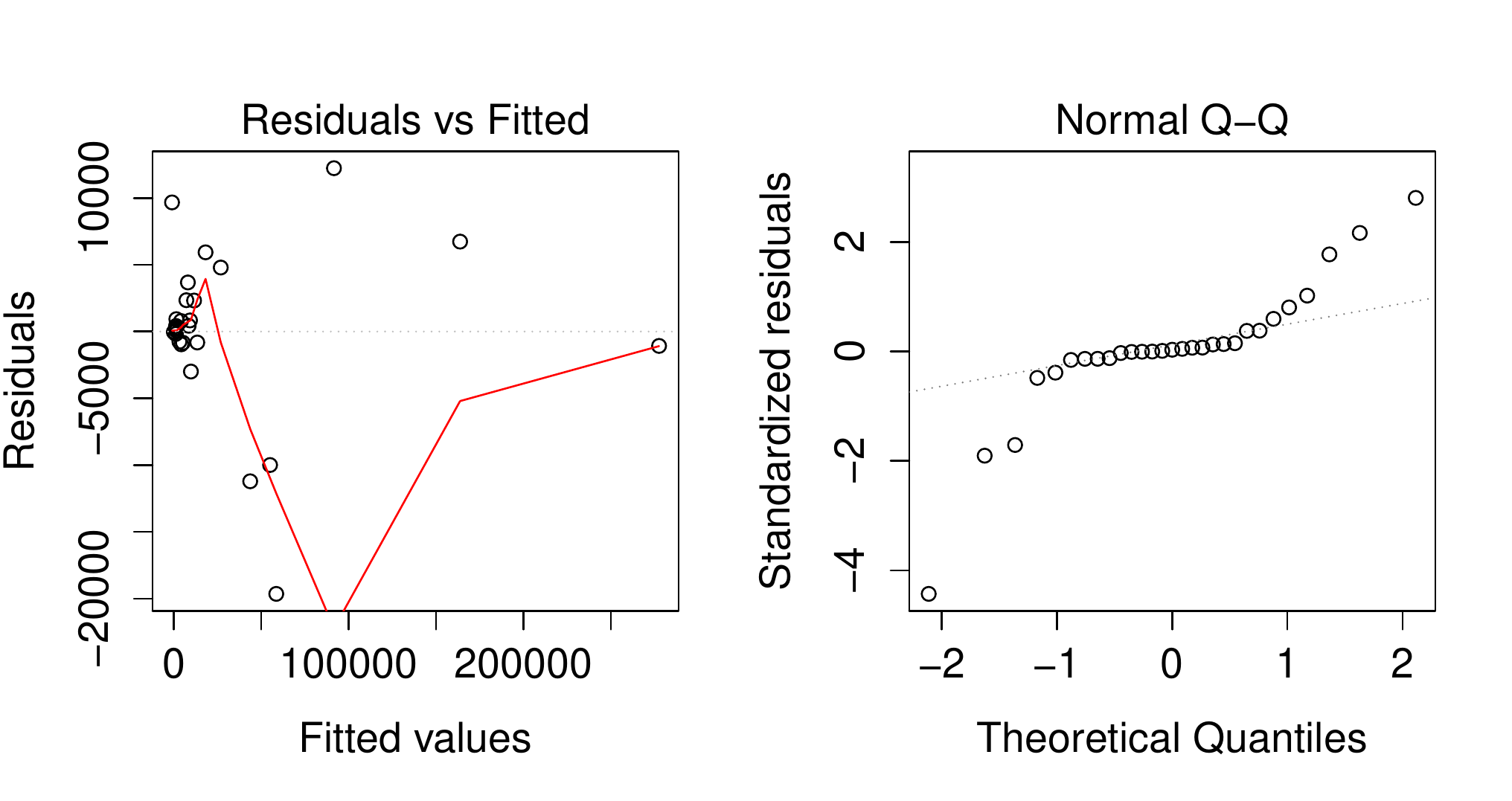}
\end{subfigure}
\begin{subfigure} {\centering b) \label{fig::DIC::N::qd::PER}}
    \centering
    \includegraphics[width=0.25\textwidth]{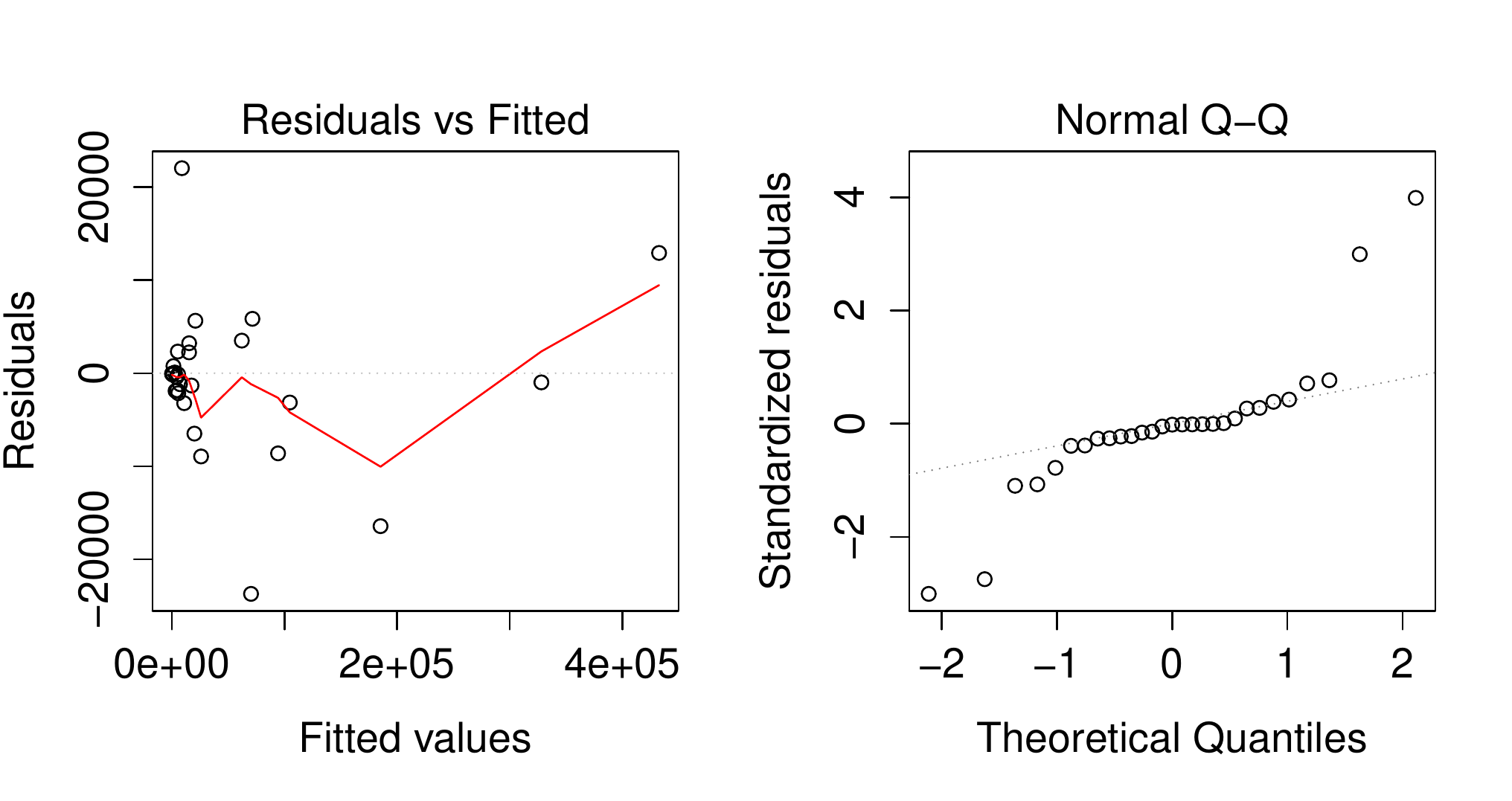}
    \end{subfigure}    
    \centering
  \begin{subfigure} {\centering c) \label{fig::MNB::P::qd::PER}}
   \includegraphics[width=0.25\textwidth]{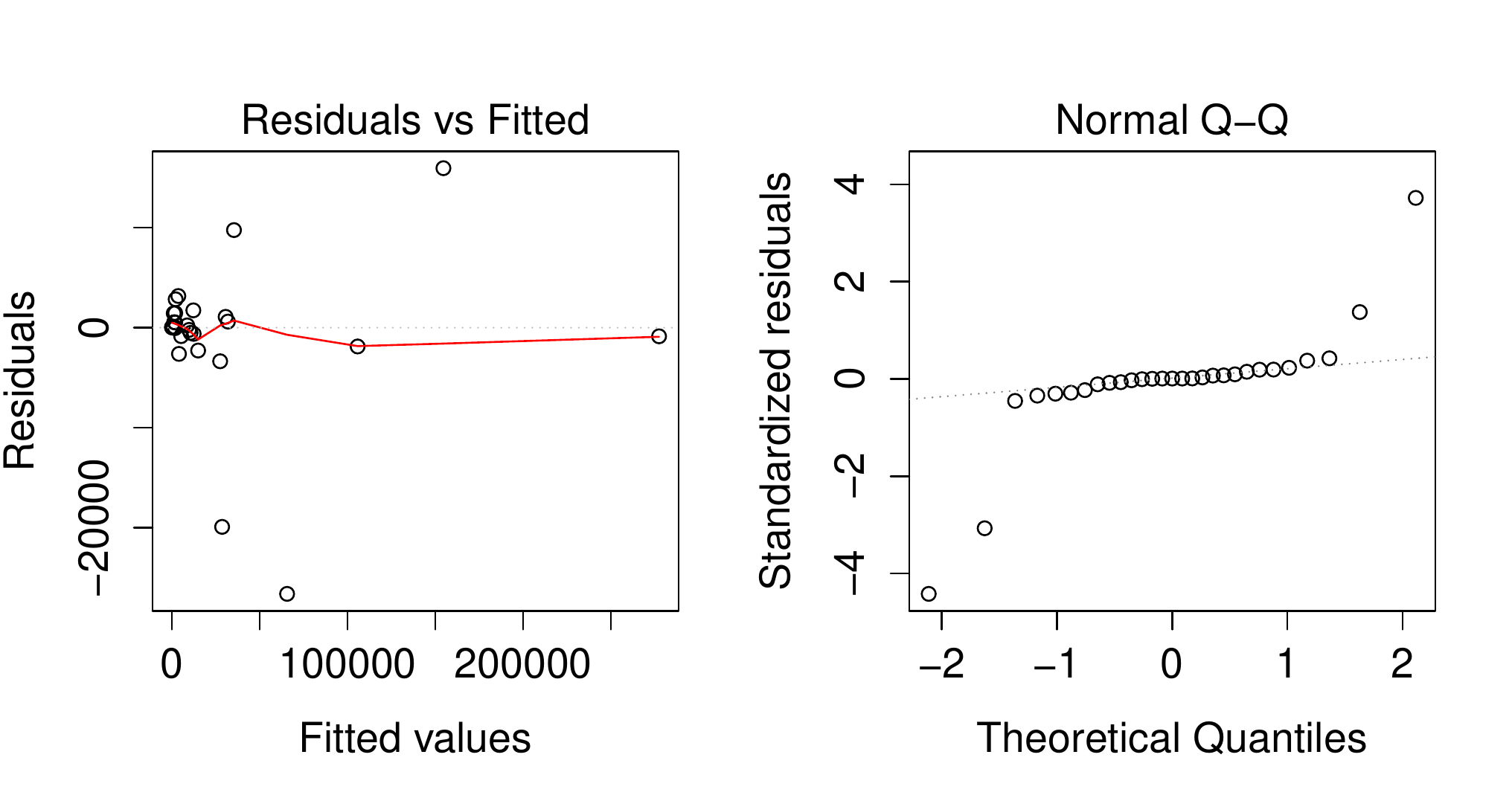}
\end{subfigure}

\begin{subfigure} {\centering d) \label{fig::MNB::N::qd::PER}}
    \centering
     \includegraphics[width=0.25\textwidth]{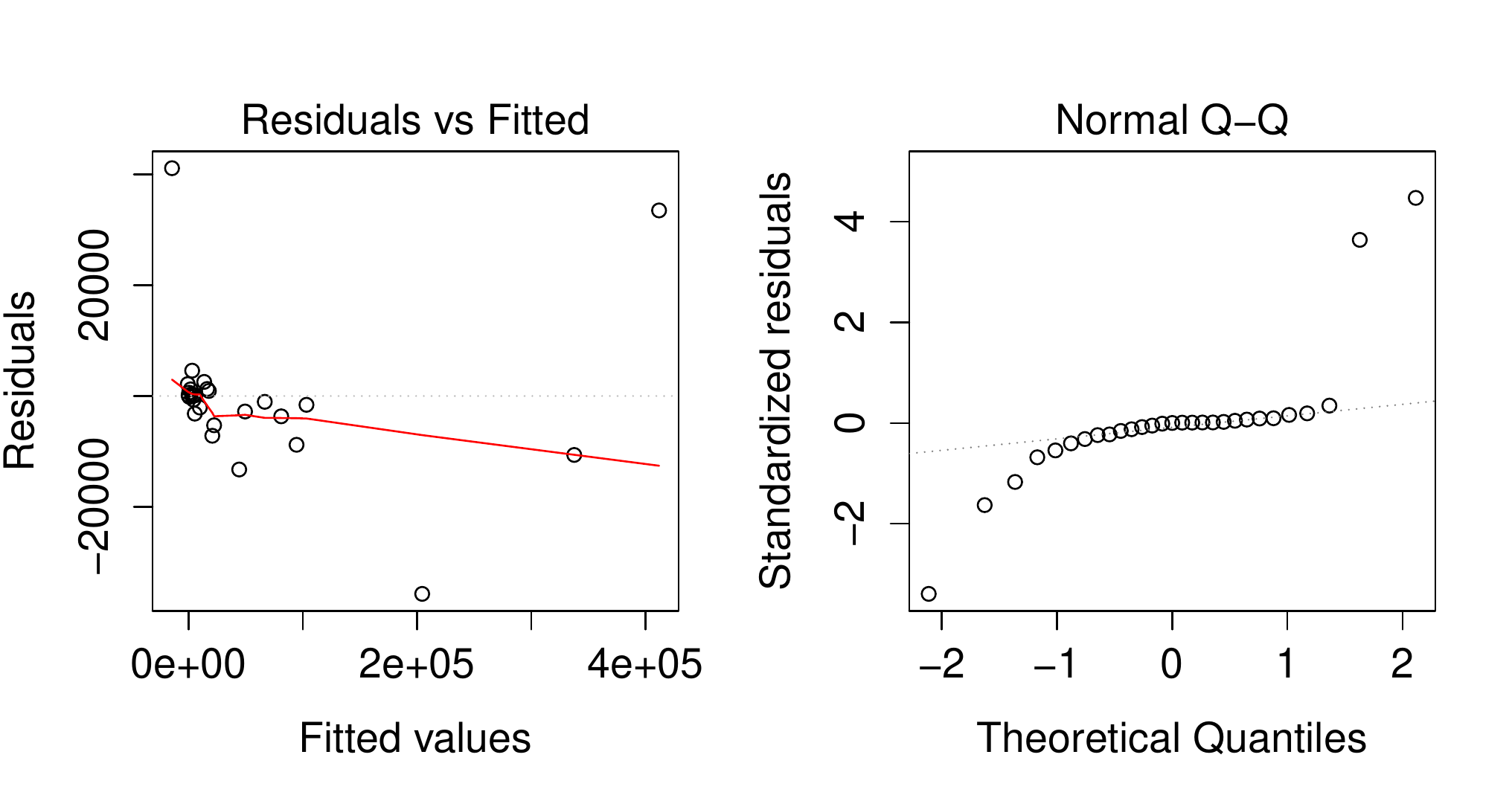}
    \end{subfigure}
\centering
  \begin{subfigure} {\centering e) \label{fig::SVM::P::qd::PER}}
   \includegraphics[width=0.25\textwidth]{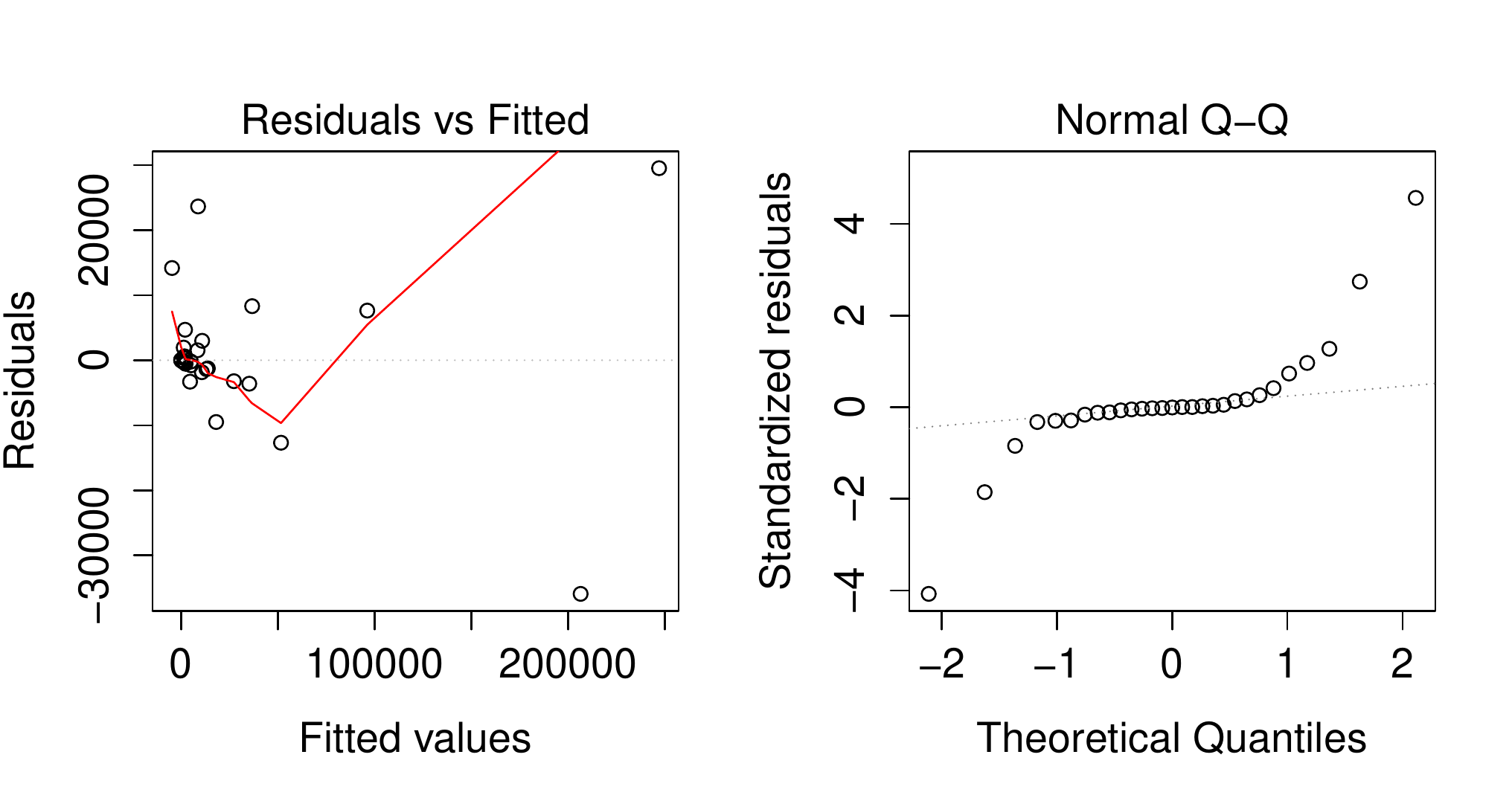}
\end{subfigure}
\begin{subfigure} {\centering f) \label{fig::SVM::N::qd::PER}}
    \centering
    \includegraphics[width=0.25\textwidth]{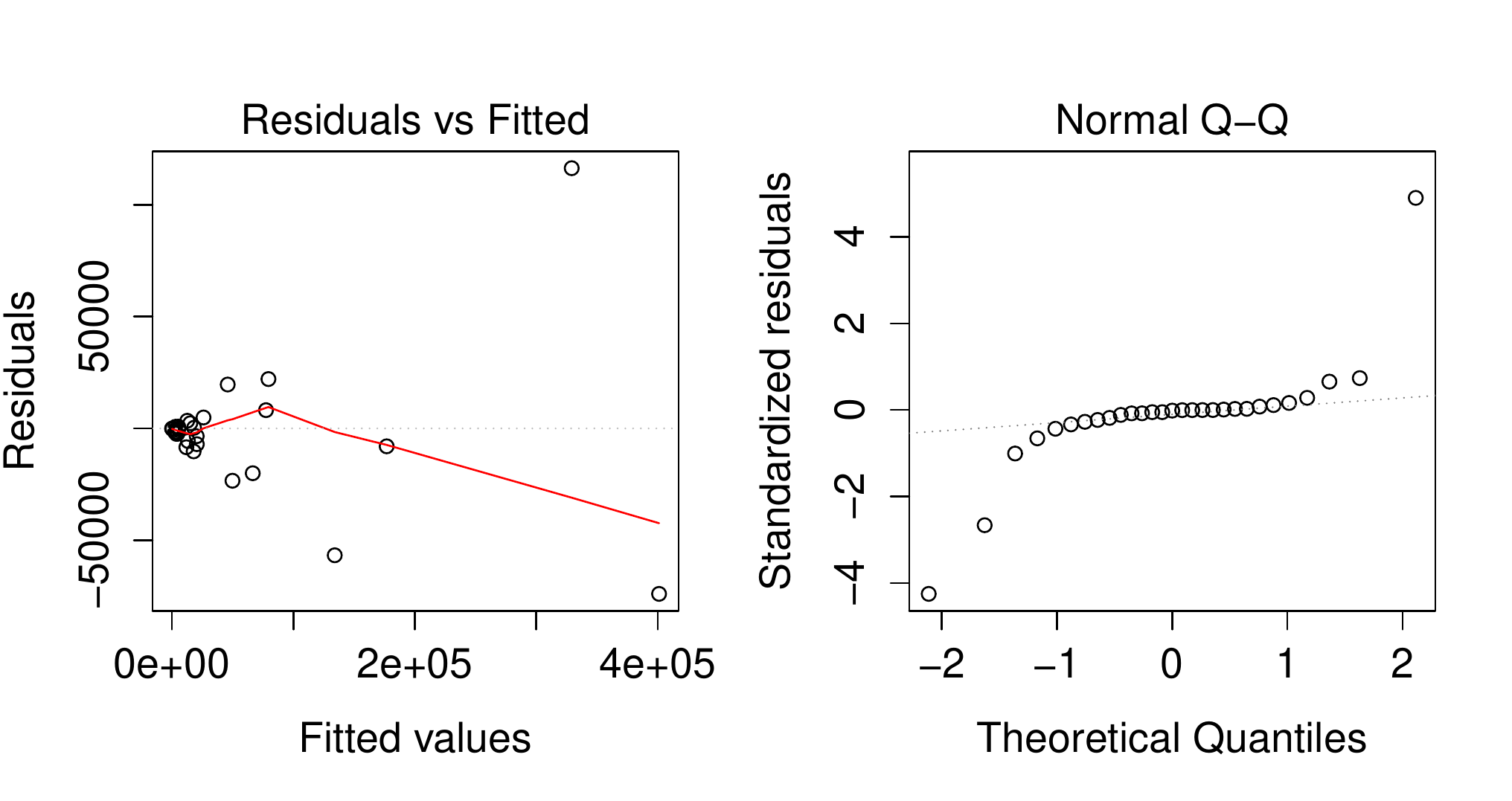}
    \end{subfigure}
    \caption{ 
    The regression parameters are learned by a query-driven experiments, that is setting one equation for each query $\query\in Q$).   \textcolor{darkred}{\bf DBM}[a) b)],
    \textcolor{blue}{\bf MNB} [c) d)], 
  \textcolor{giallino}{\bf SVM} [e) f)].
 \newline   a), c), e)  $P + M \sim   \alpha_P\cdot  \mu_P (D_\query) + \alpha_N \cdot  \mu_N (D_\query) +   \beta_P\cdot |D_\query|$ and $\query\in Q$.\newline
b), d), f) $N + M \sim   \alpha'_P\cdot  \mu_P (D_\query) + \alpha'_N \cdot  \mu_N (D_\query) +   \beta_N\cdot |D_\query|$ and $\query\in Q$.\newline
 }
\end{figure}

\begin{table}[htbp]
\caption{Pearson Correlation $\rho$ between  cumulative  measures and  observed cardinalities for a set of queries for each  classifier. The correlation is made on a leave-one-out cross validation learning observations. For each category $\category \in\{P,N\}$ the  number  of  the two categories in the collection (estimated by sampling randomly in each result set  of  the queries and manually evaluated) is correlated to the cumulative measure of the classifiers $\mu_\category$ over the  collection. We also report the confidence interval for $\rho$ at 95\% of confidence level.}

\begin{center}
\begin{tabular}{|C{2.1cm}|C{0.8cm}C{2.5cm}|C{0.8cm}C{2.5cm}|C{0.8cm}C{2.5cm}|}
\hline
classifier/$\rho$ 	& \multicolumn {2} {|c|}{DBM}	&	 \multicolumn {2} {|c|}{SVM}	&	 \multicolumn {2} {|c|}{MNB}	 \\\hline
 	& mean & interval	&	mean & interval	&	mean & interval
		 \\\hline
$\mu_P$/P (estim.)	 & 0.974 & $\rho\in[0.944, 0.988]$	
&	 0.946 &$\rho\in[0.887, 0.975]$
& 0.924 &$\rho\in[ 0.887,0.975]$ \\
$\mu_N$/N (estim.) &	0.987&	 $\rho\in[ 0.973, 0.994]$
& 0.918 &$\rho\in[0.831, 0.961]$	&	0.950  &	 $\rho\in[ 0.895,0.977]$ \\\hline
\end{tabular}
\end{center}
\label{table::correlation}
\end{table}

\begin{table}[!b]
\caption{Pearson Correlation $\rho$ between proportions with the classify and count models (CC($\category$)), cumulative models $\Phi$ and  proportions of $\category$ for the set of queries. The correlation is made on a leave-one-out cross validation learning observations. 
  We also report the confidence interval for $\rho$ at 95\% of confidence level (all p-values are $<0.05$).}

\begin{center}
\begin{tabular}{|c|c|c|c|c|c|c|c|c|c|}
\hline
 	\multicolumn{4}{|c|}{	Classify and Count					}&\multicolumn{3}{|c|}{	Query driven $\Phi$					}&\multicolumn{3}{|c|}{	Item driven $\Phi$}\\\hline
	&	DBM	&	MNB	&	SVM	&	DBM	&	MNB	&	SVM&	DBM	&	MNB	&	SVM	\\\hline
Pos $\rho$ mean	&	0.886	&	0.943	&	0.951	&	0.991	&	0.989	&	0.978	&	0.978	&	0.975	&	0.943	\\
Neg $\rho$ mean 	&	0.979	&	0.954	&	0.943	&	0.994	&	0.988	&	0.936	&	0.99	&	0.986	&	0.928	\\\hline
Pos $\rho$ Inf	&	0.769	&	0.881	&	0.896	&	0.981	&	0.977	&	0.953	&	0.954	&	0.946	&	0.88	\\
Neg $\rho$ Inf	&	0.955	&	0.902	&	0.88	&	0.987	&	0.974	&	0.866	&	0.977	&	0.97	&	0.85	\\\hline
Pos $\rho$ Sup	&	0.945	&	0.973	&	0.976	&	0.996	&	0.995	&	0.989	&	0.989	&	0.988	&	0.973	\\
Neg $\rho$ Sup	&	0.99	&	0.978	&	0.973	&	0.997	&	0.994	&	0.969	&	0.995	&	0.993	&	0.965	\\\hline
\end{tabular}
\end{center}
\label{table::correlation::2}
\end{table}

\begin{table}[htbp]
\caption{Kolmogorov-Smirnov distance $D\in[0,1]$ between the fitted and the observed distributions. The two sets of   fitted  and observed values come from the same distribution is the hypothesis under test. The  $^*$ indicates statistical significance at 95\% level of confidence. }

\begin{center}
\begin{tabular}{|c||c|c||c|c||c|c|}
\hline
& \multicolumn{6}{c|}{	Classify and Count} \\
\hline
& \multicolumn{2}{c|}{DBM} & \multicolumn{2}{c|}{MNB}& \multicolumn{2}{c|}{SVM}\\
\hline
	&	D	&	p-value & 	D	&	p-value & 	D		&	p-value		 	\\
\hline
Pos &	0.069    & 		1 $^*$	& 0.2069	&	0.5722	&	0.2759	&	0.2221			\\
Neg &	0.2414    &  	0.372		& 0.2414	&	0.3669	&	0.2069	&	0.5722			\\
\hline
\end{tabular}
\begin{tabular}{|c||c|c||c|c||c|c|||c|c||c|c||c|c|}
\hline
& \multicolumn{6}{c|}{$\Phi$ Query driven}& \multicolumn{6}{c|}{$\Phi$ Item driven}\\
\hline
& \multicolumn{2}{c|}{DBM} & \multicolumn{2}{c|}{MNB}& \multicolumn{2}{c|}{SVM}& \multicolumn{2}{c|}{DBM} & \multicolumn{2}{c|}{MNB}& \multicolumn{2}{c|}{SVM}\\
\hline
	&	D	&	p-value & 	D	&	p-value & 	D		&	p-value	&	D	&	p-value & 	D	&	p-value & 	D		&	p-value		 	\\
\hline
Pos &	0.103    & 		0.998 $^*$	& 0.172&	0.791	&	0.137	&	0.951 $^*$		&	0.620    & 		$1.5e^{-05}$	& 0.137&	0.951 $^*$	&	0.137	&	0.951 $^*$				\\
Neg &	0.137    &  	0.951 $^*$	& 0.103	&	0.998 $^*$	&	0.172	&	0.791	 &	0.413    &  	0.013		& 0.137	&	0.951 $^*$	&	0.103	&	0.998 $^*$			\\
\hline
\end{tabular}
\end{center}
\label{table::correlation::ks}
\end{table}

\section{Linear Regression Model}
Cumulative quantification models  have two sets of parameters: the parameters of the classifier (e.g. the support vectors and the parameter $b$ for SVM, $L_\category$, $L$, $f_{i,\category}$ and $\pi_i$ for DBM and MNB) and the parameters of the learning model that correlates the cumulative  measure of the classifier to the counting measure of  categories size. We expect that the relation between the two measures is expressed by a linear correlation, therefore we choose the linear regression as  natural  learning model to express such a correlation (see Table \ref{table::correlation}). 

We assume that we have already learned the classifier generating thus a cumulative measure $\mu_\category$ for each category $\category \in {\cal C}$. We now consider a second learning model $\Phi$ that learns how to predict category size and proportions from the cumulative measures $\mu_\category$. 
The validation data set $\Omega$ is made of about 3 million tweets. In order to validate the quantification model $\Phi$ we exclude all relevant and evaluated tweets of a query $\query$ both to learn   $\mu$ and  to predict the category sizes for the retrieval set of that query.
The set of evaluated tweets $V$ is extracted from  a proper subset of $\Omega$: $V=\cup_{\query\in Q}\displaystyle V_\query \textrm { with }     V_\query\subset R_\query$
where  $R_\query$ is the set of relevant  documents that  fall into 5 mutually exclusive categories: \[R_\query =\left [ P_\query\cup N_\query \cup M_\query\cup X_\query\cup O_\query \right]\]
The category rates  are estimated  using the set of  evaluated tweets $V_\query$, i.e.:
$\hat\category^\%_\query = \frac{R_\query\cap \category_\query\cap V_\query}{R_\query\cap V_\query}= \frac{ \category_\query\cap V_\query}{V_\query}$, with $\category\in \{P,N,M,X,O\}$. The actual values $c^\%_\query=\frac{ \category_\query\cap R_\query}{R_\query}$ fall into a confidence interval, i.e. $c^\%_\query = \hat c^\%_\query\pm \epsilon$.

The linear regression models of the query-driven approach are learned by using the values   $\mu_P (D_\query)$ and $\mu_N (D_\query)$ computed on the entire result set $D_\query$:
\begin{eqnarray}
\hat P^\%_\query +M^\%_\query\sim \alpha_P \cdot  \frac{\mu_P (D_\query) }{\sum_{\category\in\{P,N\}} \mu_\category (D_\query)}+  \alpha_N \cdot  \frac{\mu_P (D_\query)}{\sum_{\category\in\{P,N\}}\mu_\category (D_\query)} \textrm { s.t.  } \query\in Q 
\label{eq::lr::pos::queryDriven}\\
\hat N_\query +M^\%_\query\sim  \alpha'_P \cdot  \frac{\mu_P (D_\query)}{\sum_{\category\in\{P,N\}}\mu_\category (D_\query)} +  \alpha'_N \cdot  \frac{\mu_P (D_\query)}{\sum_{\category\in\{P,N\}}\mu_\category (D_\query)} \textrm { s.t.  } 
\query\in Q \label{eq::lr::neg::queryDriven}
\end{eqnarray}
The number of  positive documents $P$ is estimated by $\hat P_\query = |D_\query| \cdot \hat P^\%_\query$, and similarly with the set of negative documents $\hat N_\query = |D_\query| \cdot \hat N^\%_\query$, where $\hat P^\%_\query$ and $\hat N^\%_\query$ are evaluated on the set $V_\query$, and $D_\query$ is the retrieval set for the query $\query$.

Differently,  the item-driven approach learns the  regression parameters by means of the following set of equations:
\begin{eqnarray}
p(\vec x) \sim \alpha_P \cdot  \mu_P (\vec x) +  \alpha_N \cdot  \mu_P (\vec x) \textrm { if } \vec x\in P\cup N 
\label{eq::lr::pos::itemDriven}\\
 n(\vec x) \sim  \alpha'_P \cdot  \mu_P (\vec x) +  \alpha'_N \cdot  \mu_P (\vec x)  \textrm { if } \vec x\in P\cup N  
\label{eq::lr::neg::itemDriven}
\end{eqnarray}
where $p(\vec x)$ is equal to $1$ if the document $\vec x$ is positive and $0$ if it is negative; while, $n(\vec x)$ is equal to $1$ if the documents $\vec x$ is negative and zero if it is positive.

Once the regression parameters are learned with a Leave-One-Out cross validation, for each query the sizes of the positive set and  the negative set are estimated with a cumulative approach as follows:
\begin{eqnarray}
P_\query = \alpha_P\cdot  \mu_P (D_\query) +  \alpha_N \cdot  \mu_N (D_\query)
\label{eq::size::pos::itemDriven}
\\
N_\query = \alpha'_P \cdot  \mu_P (D_\query) +  \alpha'_N \cdot  \mu_N (D_\query)
\label{eq::size::neg::itemDriven}
\end{eqnarray}
where $\query\in Q$, and $D_\query$ is the result set computed using data set $\Omega$. Finally, the percentage for each category $c$ and for each query $\query$ are 
\begin{eqnarray}
  P^{\%}_\query = \frac{P_\query}{ P_\query + N_\query}
\label{eq::per::pos::itemDriven}
& \hskip 2cm
N^{\%}_\query = \frac{N_\query}{ P_\query + N_\query}.
\label{eq::per::neg::itemDriven}
\end{eqnarray}

Thanks to the additivity property of $\mu_\category$ we also note that Equations \ref{eq::size::pos::itemDriven} and \ref{eq::size::neg::itemDriven} are indeed the sum of \ref{eq::lr::pos::itemDriven} and \ref{eq::lr::neg::itemDriven} respectively.

\section{Experiments}\label{sec::experiments}

The evaluation measures for quantification models are  based on a   value  aggregating the pairs with the observed and the predicted values  for each category $\{(y,\hat y)\}_{\category\in {\cal C}}$. The main problem of sentiment quantification  in a retrieval scenario  is that  there is a very high variability of sentiment category priors with real queries.  Such a variance  thus affects the performance of any classifier, and we therefore need to address specifically  how to measure quantification performance with a set of queries and category observations,  $\{(y,\hat y)\}_{\category\in {\cal C}, \query\in Q}$.

 However, the available collections for training and test the classifiers are of the order of few thousand  of evaluated tweets, often on a single topic (OMD, HCR,GASP,WAB) or on a few similar topics (Sanders).   SemEval contains many queries (about 180) but with very small validation and retrieval sets per query. There are about 3,000 tweets containing a 11\% of negatives and 34\% of positives with an average of 2 (5) negative (positive) tweets per  query \cite{rosenthal:2015:SemEval}.

 In addition, evaluation measures for quantification models have some drawbacks when applied to retrieval and quantification accuracy.
  One of evaluation measure used for quantification is   the mean of the residuals (absolute errors, AE), that in our case would be the mean of the values  $y-\hat y$, one for  query and each category. 
AE is biased by the queries that possess a very large result set. To overcome this problem,  one can use the mean of the rates  (RAE) instead of the mean of absolute values. The bias with RAE is that it may  be very low  with very small category. To avoid the size of the result set, one can use the mean of divergence measure between the category distributions, e.g. the Kullback-Leibler divergence (KLD). Since   KLD is  unbounded, one can alternatively apply the logistic function to this divergence to normalize it,  (NKLD) \cite{GaoSebastiani:ASONAM2015}.
A comparison between models may be conducted with a statistical test, e.g. Wilcoxon, to show whether one model is statistically better than a second one\cite {GonzaLez-Castro:2013}.

Using  a very different approach,   statistical analysis studies the distribution of the residuals $\{(y,\hat y)\}_{\category\in {\cal C}, \query\in Q}$ 
  with respect to their principal moments, in order to validate how much the learning model fits the data within a given confidence level (the margin of allowed error). The first advantage  of using statistical analysis for quantification is that  we can assess how good is a model independently from other learning models. Then we may  always compare new models according only  to the fitting parameters and their values.
Second, the number of queries used to validate the fitting becomes an important parameter to pass the significance test of the goodness of the fit.
 Third, we can distinguish  possible outliers of the model, and we may correct hypothesis and improve models. For example, according to the normal Q-Q plot one should expect that all observations pairs should lie around a straight line.  
 Therefore we  use  a validation collection \cite{DART2014} possessing the following properties:
 \begin{itemize}
\item The collection should be large with a number of test queries $Q$.
\item The result sets $D_\query$ of $Q$ may  vary largely from query to query.
\item The size $y=n^\query_\category$ of a category in each  result set is estimated by $\hat y=\hat\category^\query$ such that $|\frac{n^\query_\category}{D_\query} -\frac{\hat\category^\query}{D_\query} |\leq\epsilon$ with a confidence level of $95\%$.
\end{itemize}
 a)  To assess the scalability of the cumulative hypothesis, we need to build a large collection of tweets using some generic terms as seeds and then running a number $Q$ of queries. 
 b)  Differently from TREC collections where evaluation is focused on precision of rankings and it is thus performed by pooling  the topmost (pseudo-relevant) documents   from each running system, we  here need to estimate the size of the set of  relevant    documents into categories.  Therefore  we estimate by  sampling {\em randomly}   from the result set, and assessing a {\em sufficient}  (see condition c  below) number $V_\query$  of documents with respect to six mutually exclusive categories: P  for only positive, N for only Negative, X for neutral, M for mixed polarity, O for other, and the last category  containing the rest of all non-relevant documents. All other five sentiment categories thus co-occur with the relevance event. \\ 
  c)  We then need to decide how many documents $V_\query$ to assess for each query in order to have a statistical significant estimate of category size within a given confidence interval. This confidence interval depends on the fixed confidence level, e.g. 95\%.  We know that the size  of the validation set  $V_\query$ must be of the order of $\frac{\sqrt {D_\query}}{\sigma}$ where $\sigma$ is the standard deviation of the category size.\\
 d) To avoid the query bias, we  use the Leave-One-Out cross-validation to learn the classifier without the query  $\query\in Q$ and obtain $\{(y,\hat y)\}_{\category\in {\cal C}}$ of this query.\\
e) Finally, we study the  distribution of the residuals to assess the precision of estimates for the categories.
 The collection contains 29 queries, with a median of 30,741 and a mean of 105,564 retrieved documents for a total of about 3 million retrieved documents. The  rate of positive documents has mean  20.8\%, the maximum  44.8\% and the minimum  5.9\%, while the mean rate of negative documents is  36.9\%, the maximum  66.8\% and the minimum  11.2\%.

\begin{figure}[htpb]
    \centering
    \begin{subfigure}{a)
    \includegraphics[width=0.25\textwidth]{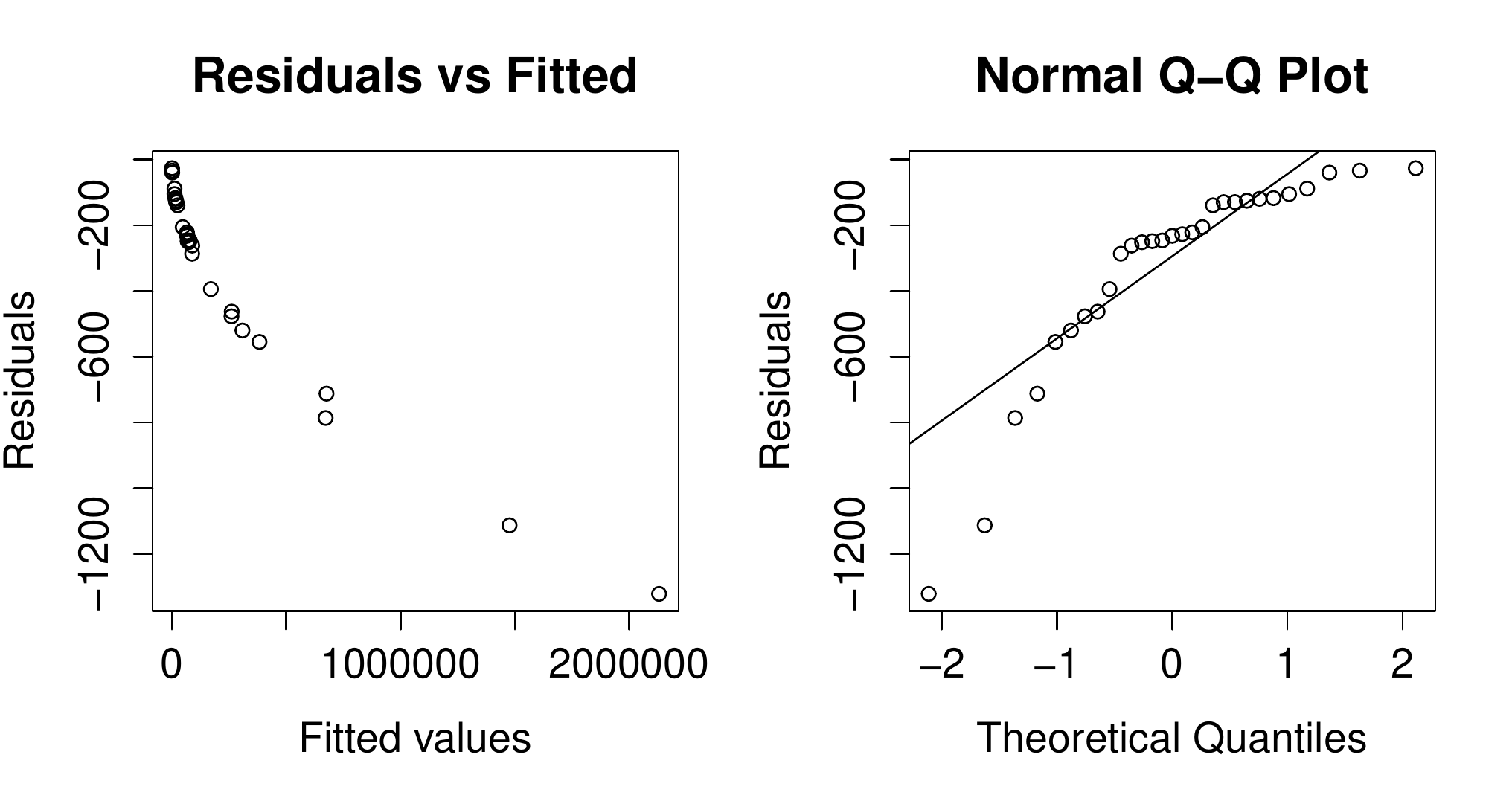}}
    \end{subfigure}
    \centering
    \begin{subfigure}{b)
    \includegraphics[width=0.25\textwidth]{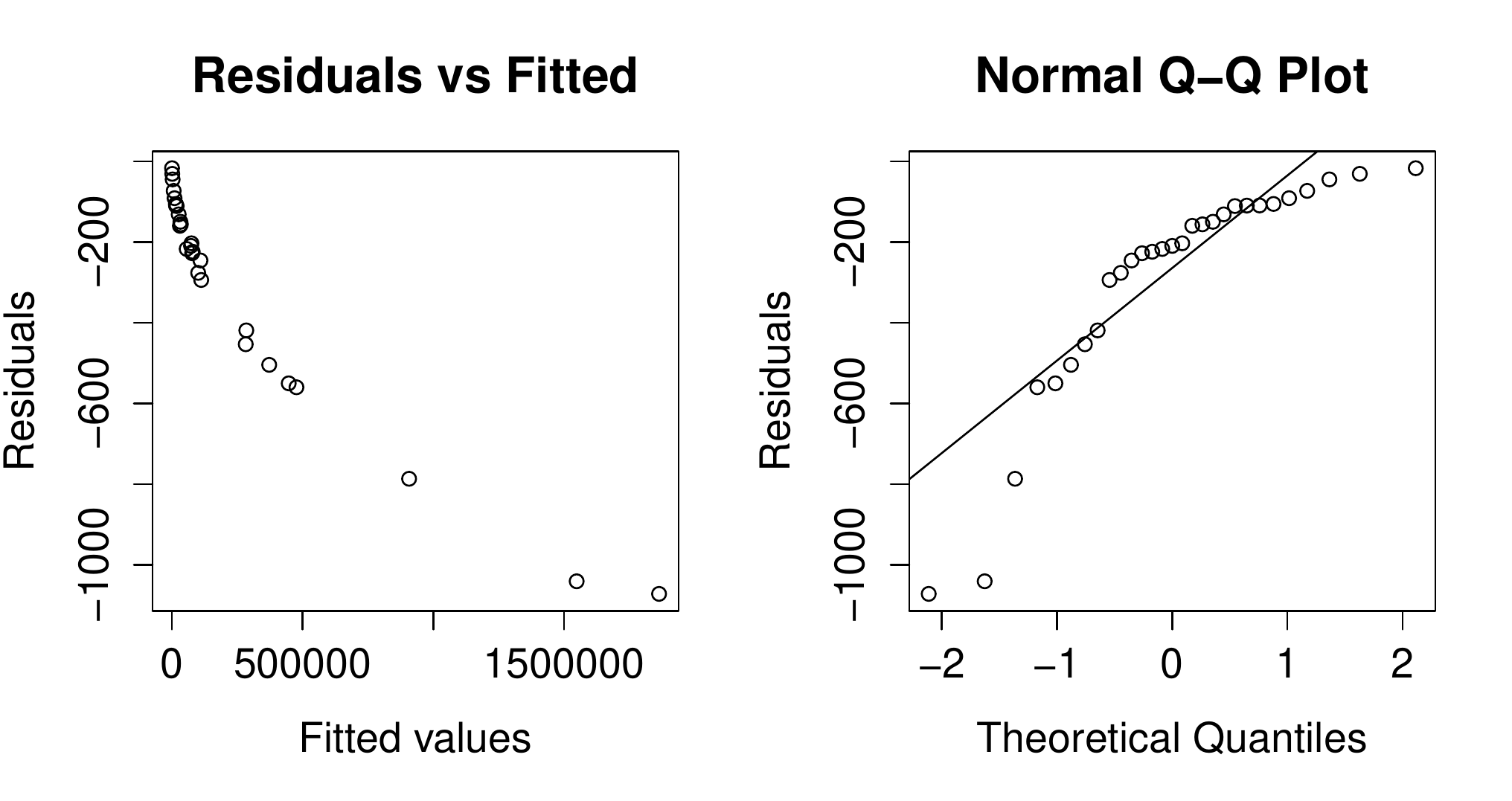}}
    \end{subfigure}
    \centering
    \begin{subfigure}{c)
    \includegraphics[width=0.25\textwidth]{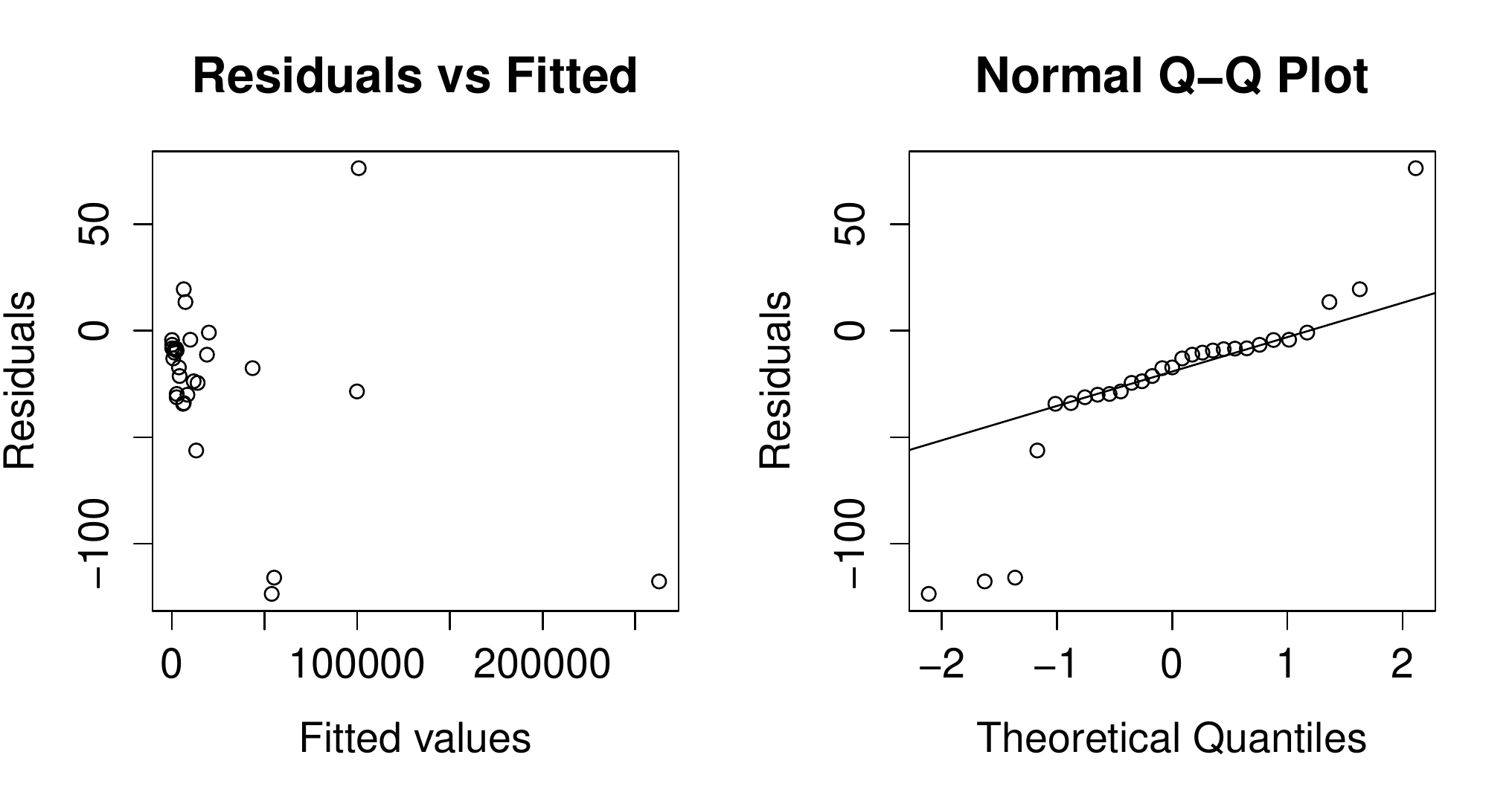}}
    \end{subfigure}
     
    \centering
    \begin{subfigure}{d)
    \includegraphics[width=0.25\textwidth]{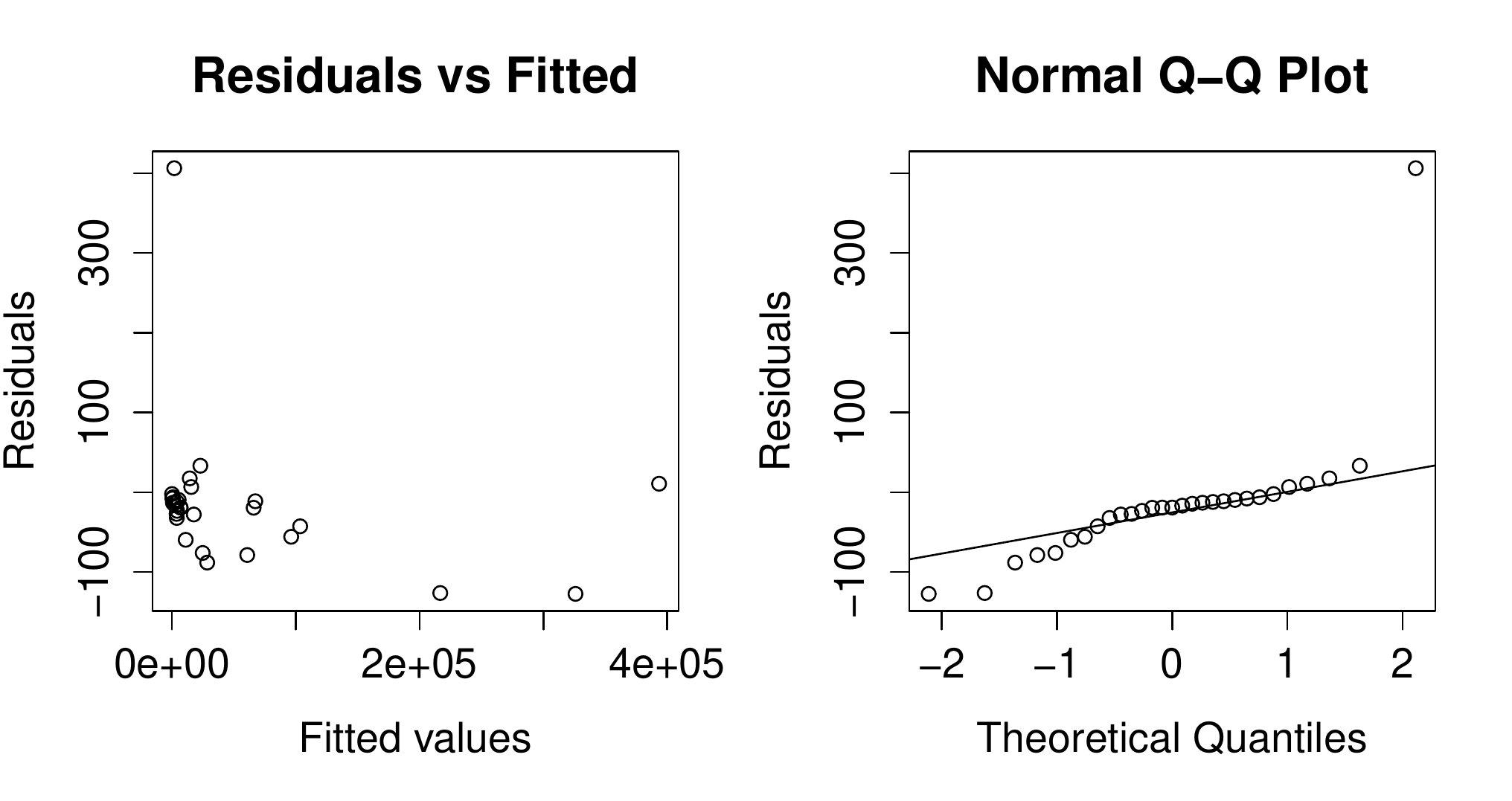}}
    \end{subfigure}
    \centering
    \begin{subfigure}{e)
    \includegraphics[width=0.25\textwidth]{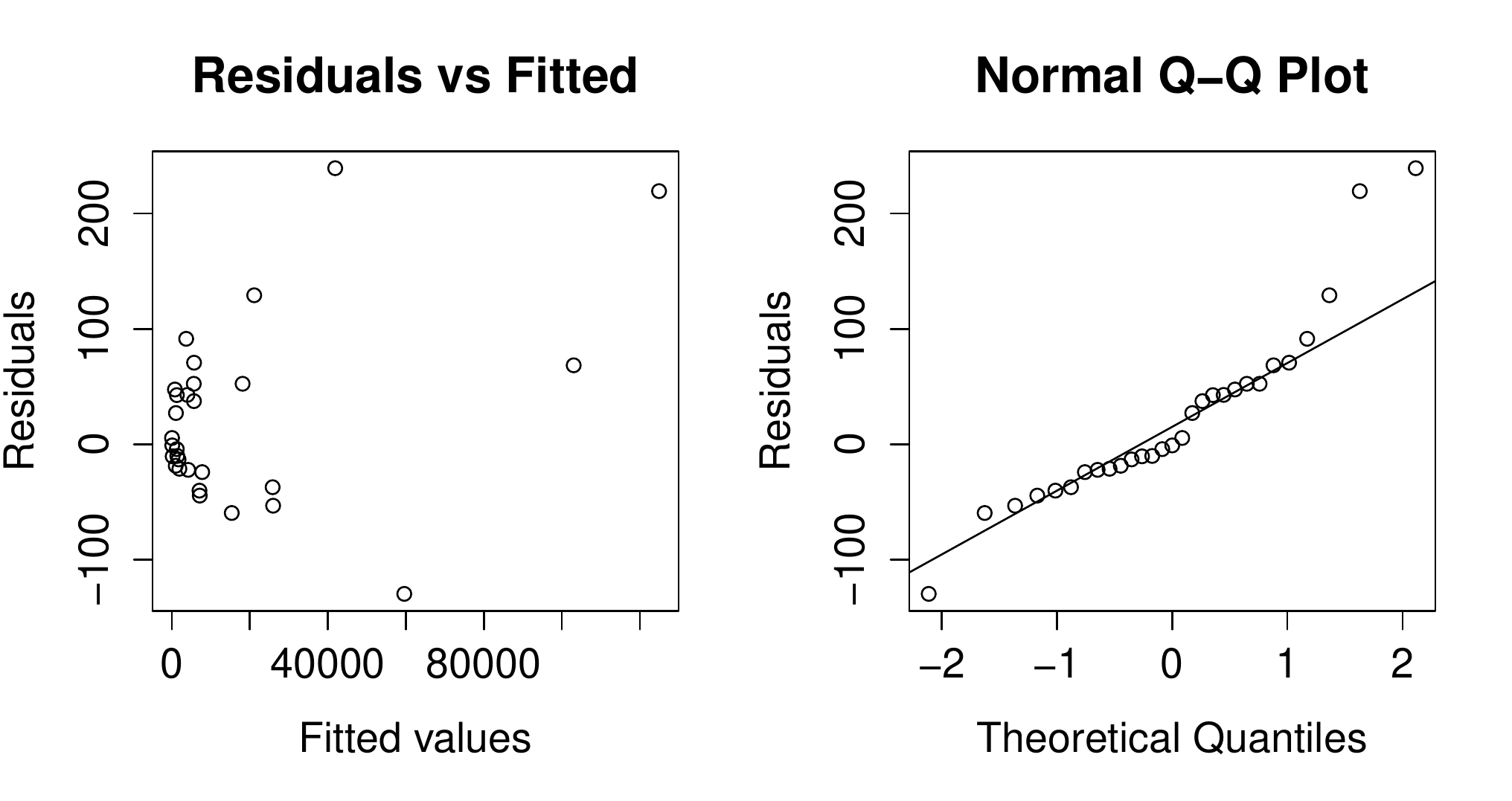}}
    \end{subfigure}
    \centering
    \begin{subfigure}{f)
    \includegraphics[width=0.25\textwidth]{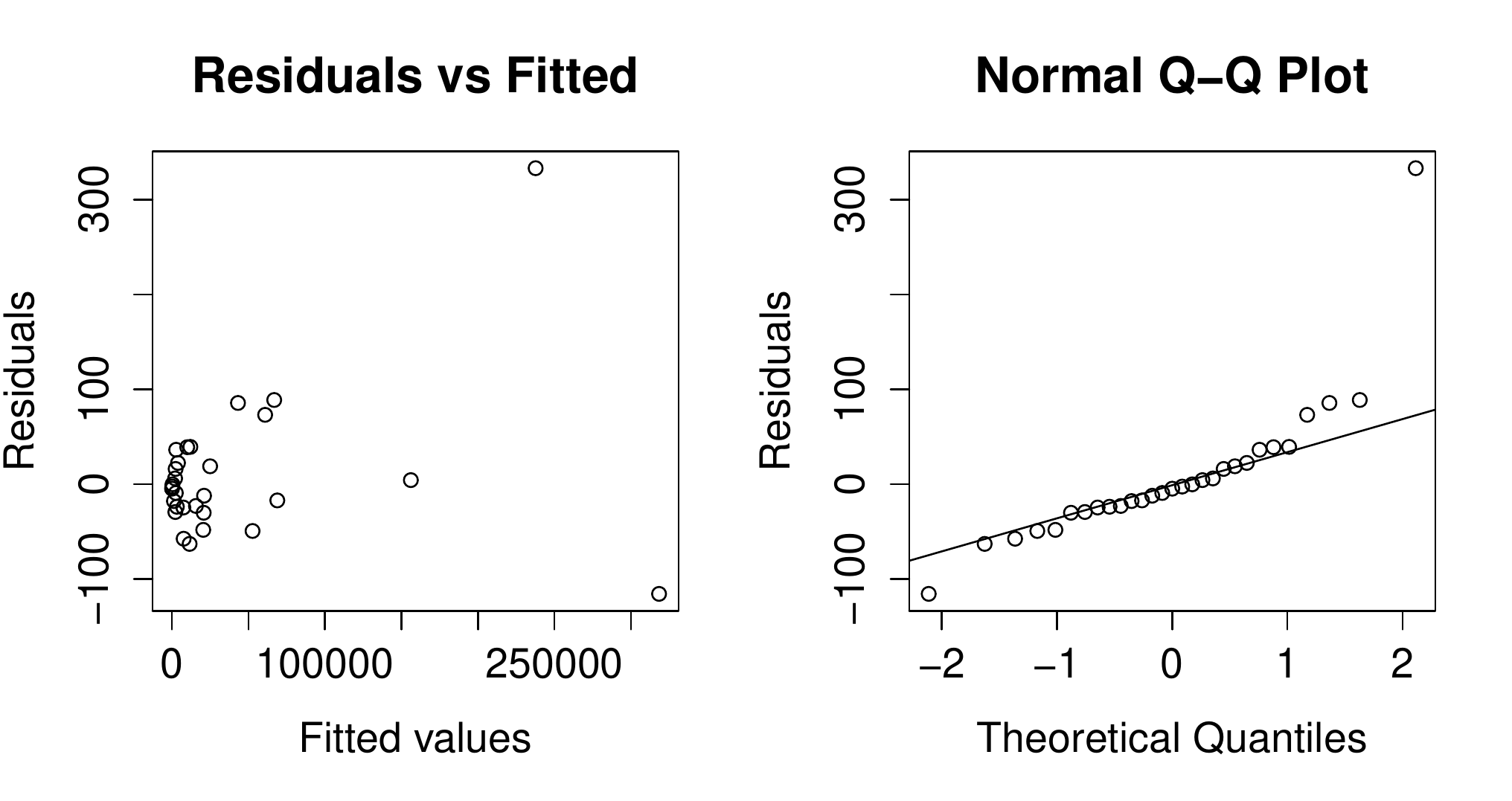}}
    \end{subfigure}

     \caption{$\Phi$ with the \textcolor{darkgreen}{\bf item driven} approach. 
  \textcolor{darkred}{\bf DBM}[a) b)],
    \textcolor{blue}{\bf MNB} [c) d)], 
  \textcolor{giallino}{\bf SVM} [e) f)].
   Positive fitted values [a) c) e)],
     Negative  fitted values [b) d) f)].
}
\end{figure}

\begin{figure}[htpb]
    \centering
    \begin{subfigure}{a)
    \includegraphics[width=0.25\textwidth]{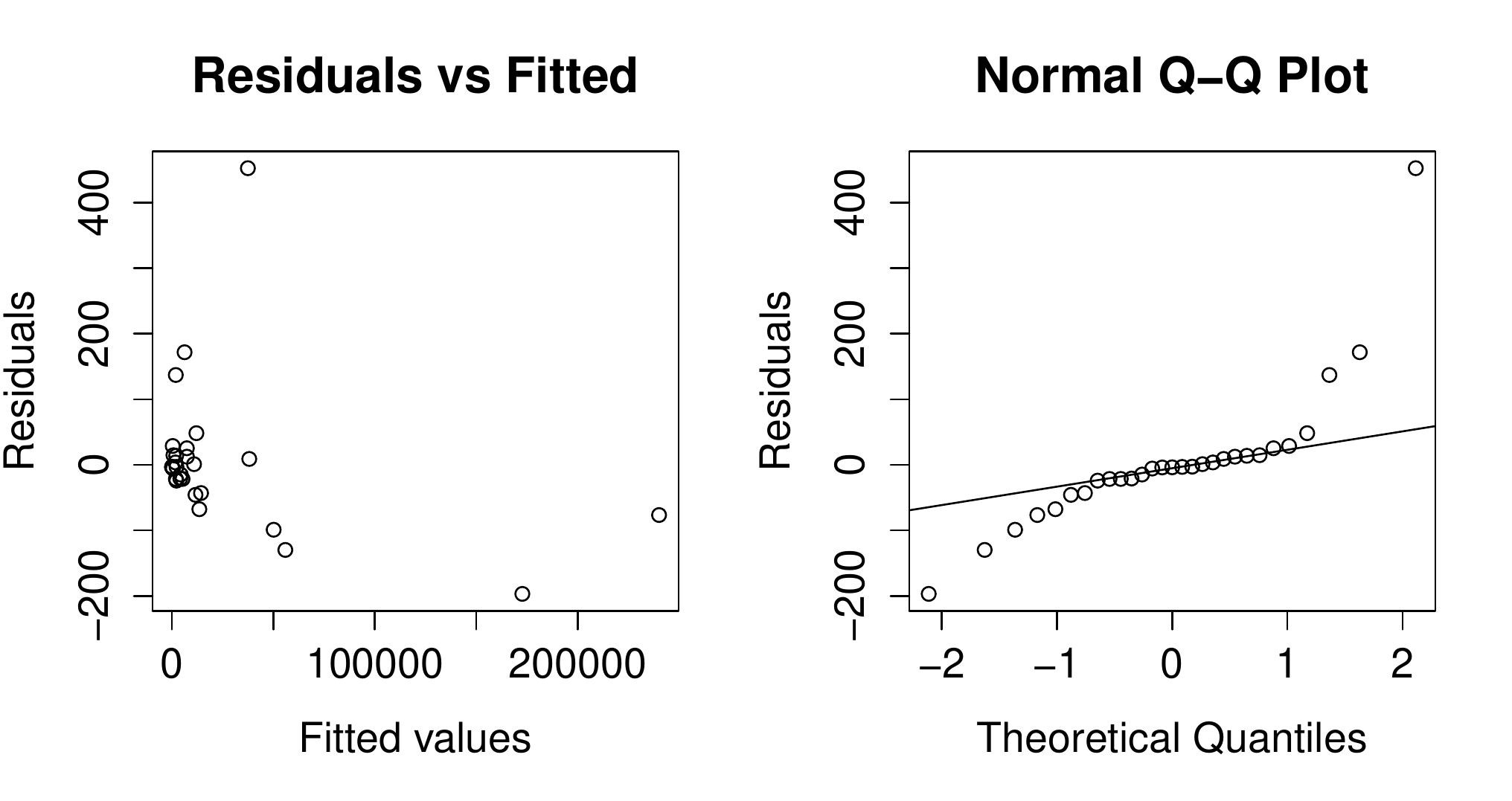}}
    \end{subfigure}
    \centering
    \begin{subfigure}{b)
    \includegraphics[width=0.25\textwidth]{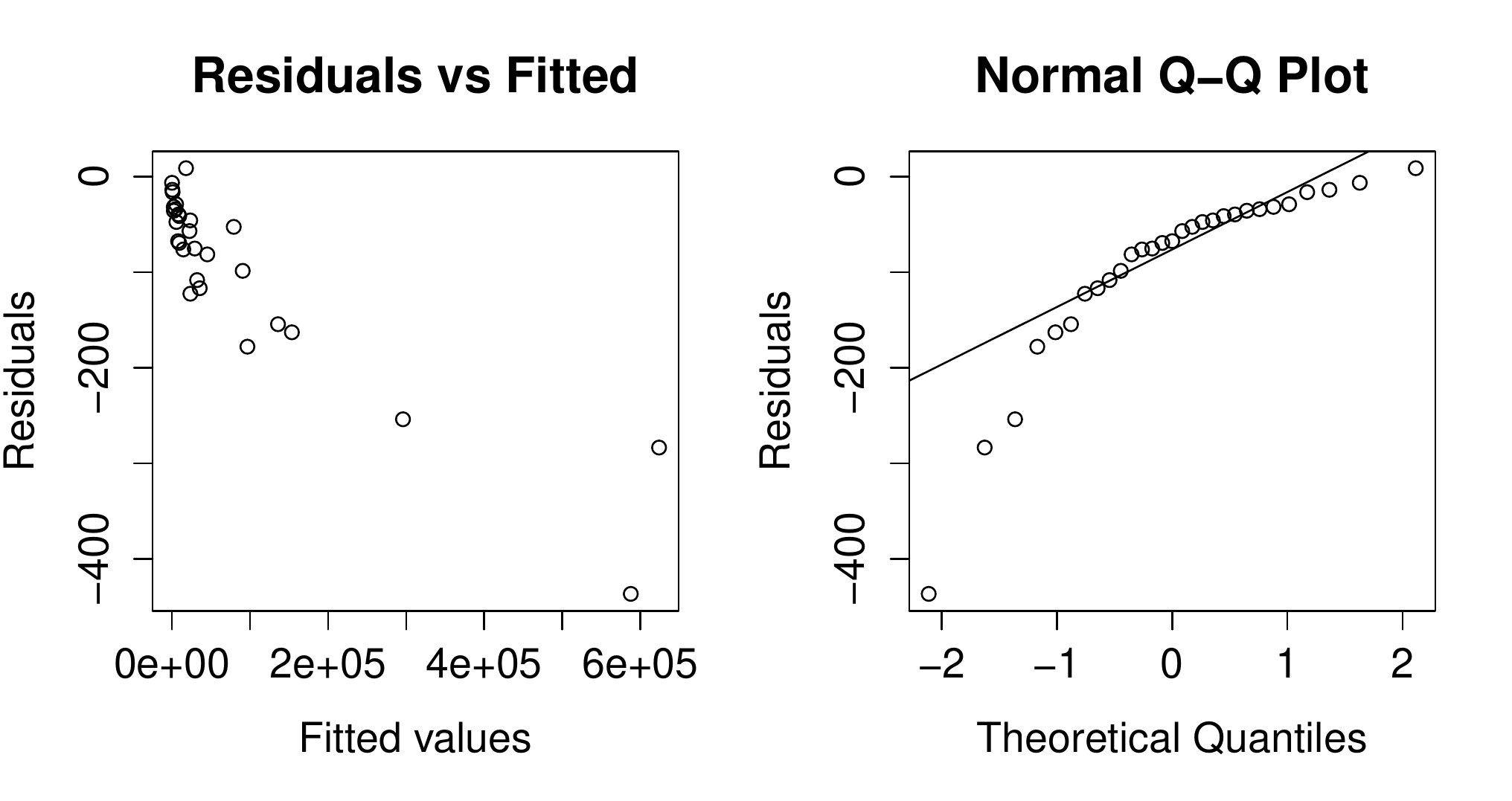}}
    \end{subfigure}
    \centering
    \begin{subfigure}{c)
    \includegraphics[width=0.25\textwidth]{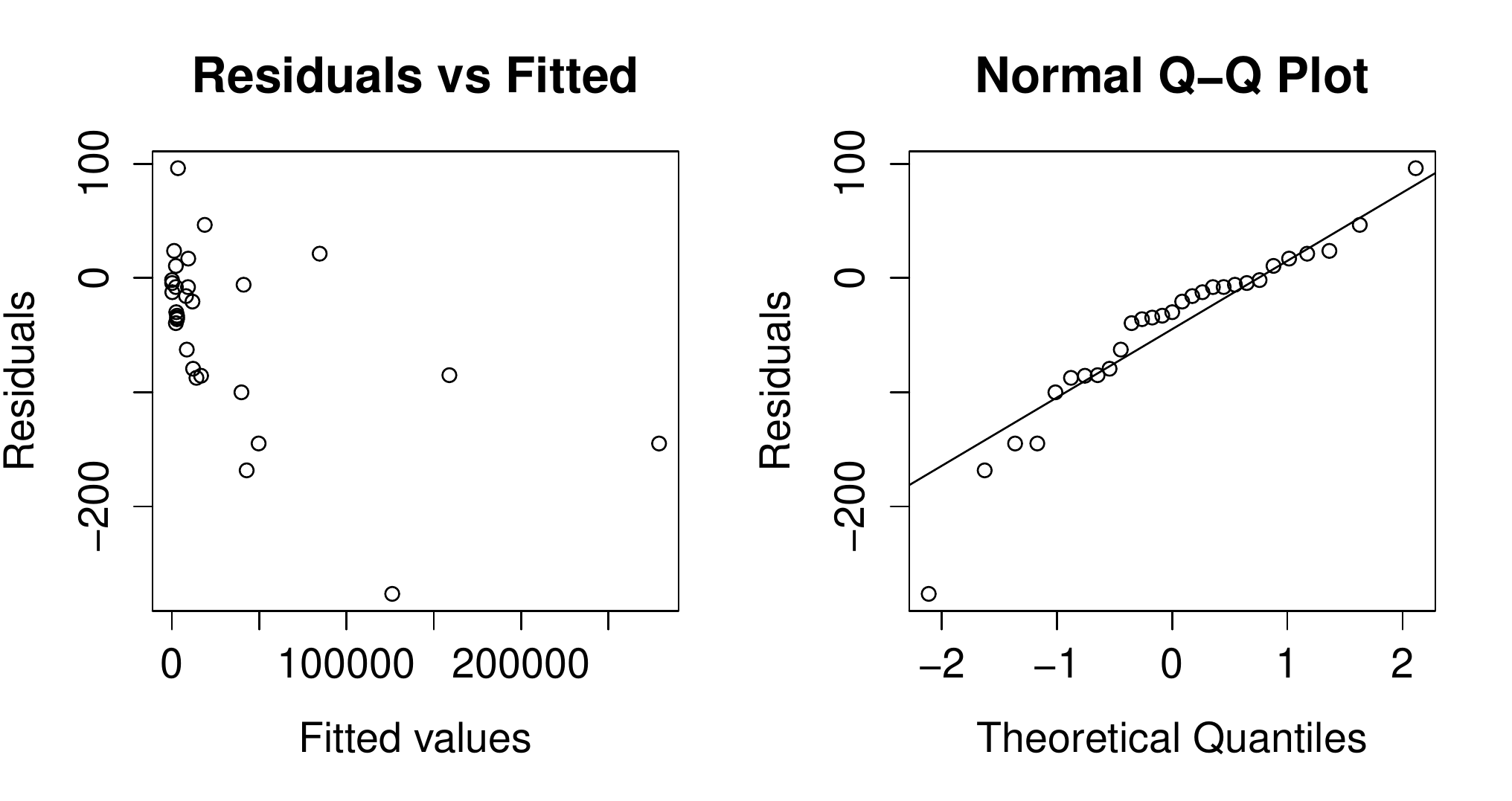}}
    \end{subfigure}

    \centering
    \begin{subfigure}{d)
    \includegraphics[width=0.25\textwidth]{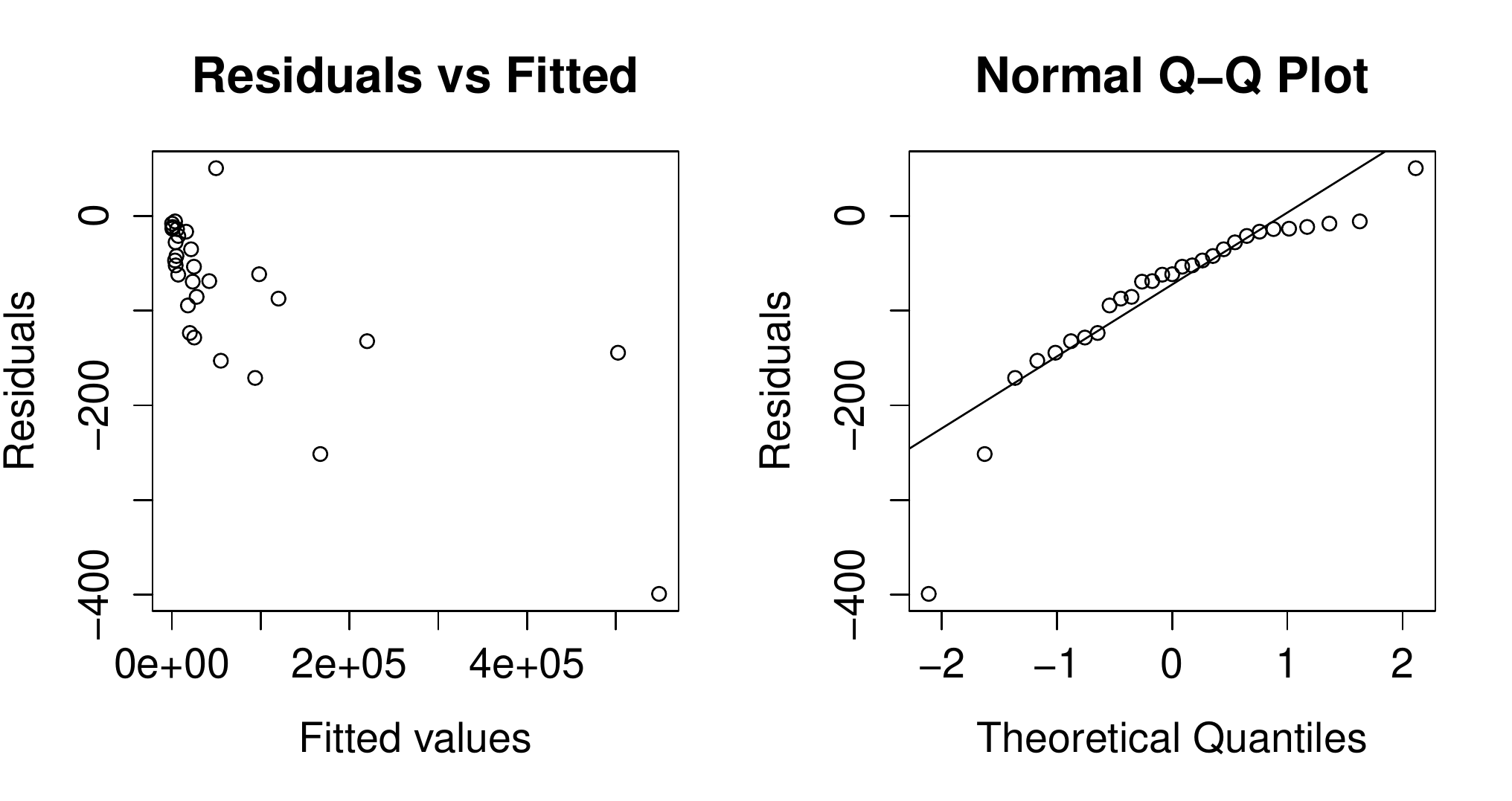}}
    \end{subfigure}
    \centering
    \begin{subfigure}{e)
    \includegraphics[width=0.25\textwidth]{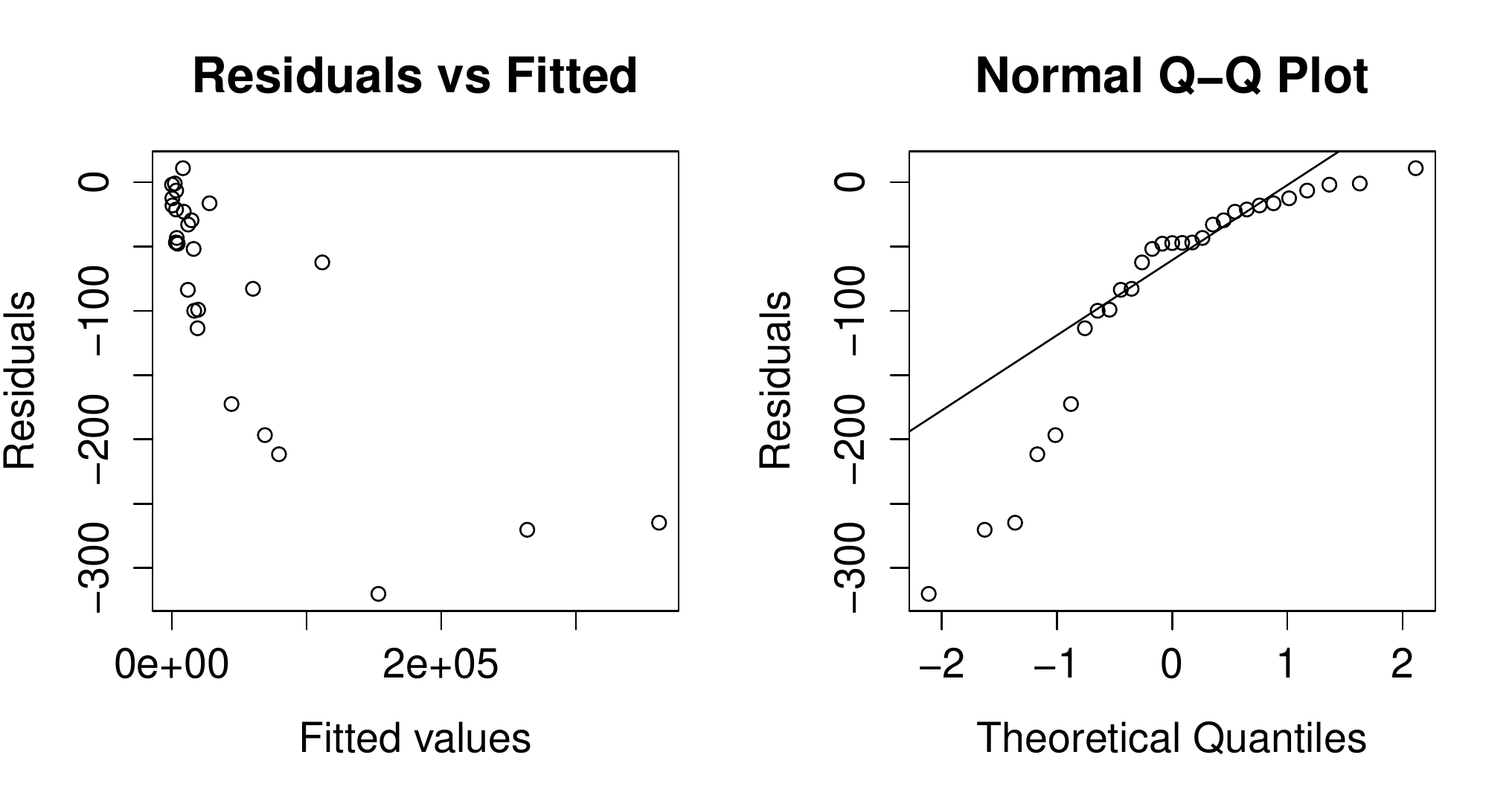}}
    \end{subfigure}
    \centering
    \begin{subfigure}{f)
    \includegraphics[width=0.25\textwidth]{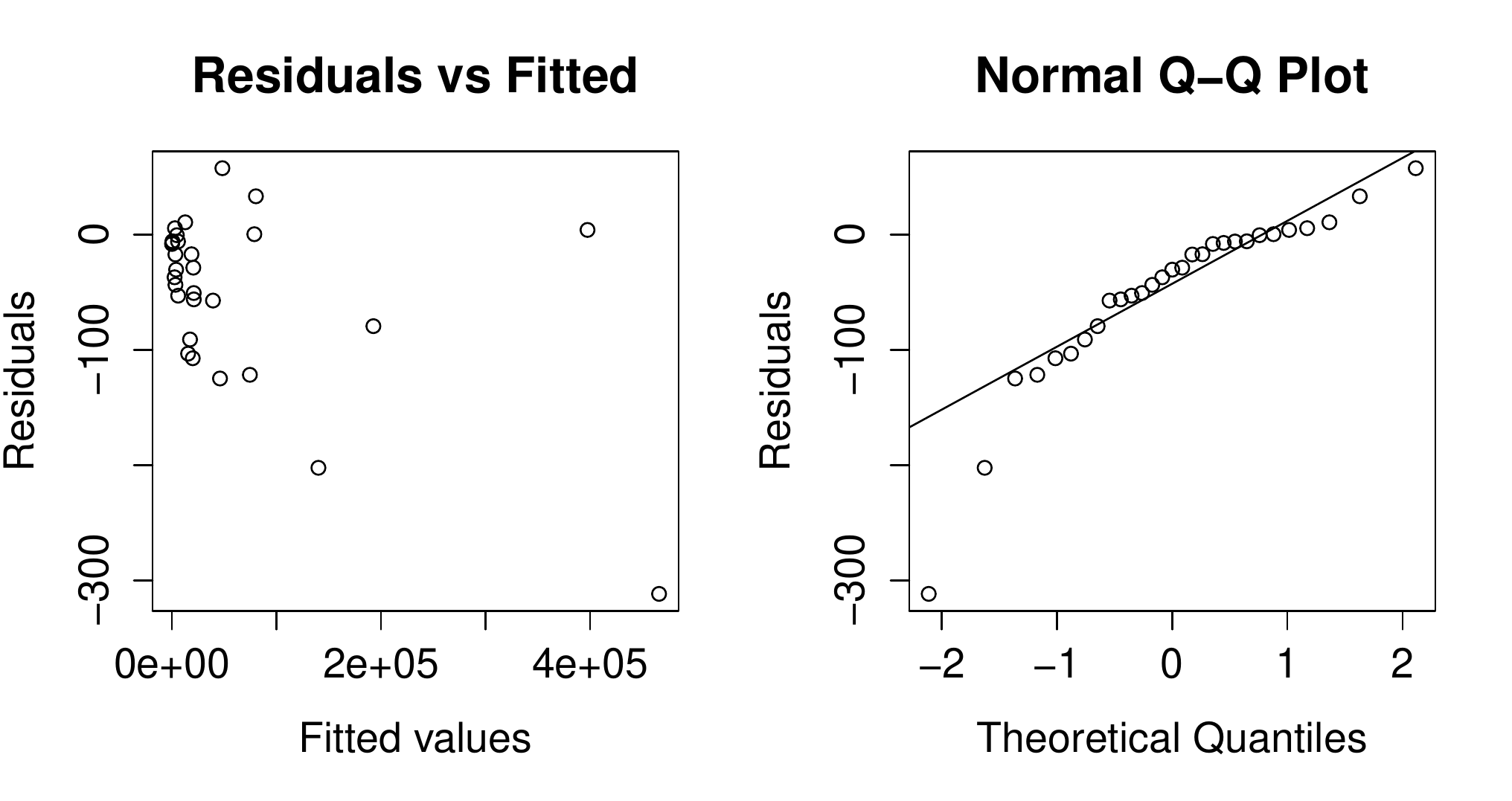}}
    \end{subfigure}
    \caption{ Classify and Count. 
  \textcolor{darkred}{\bf DBM}[a) b)],
    \textcolor{blue}{\bf MNB} [c) d)], 
  \textcolor{giallino}{\bf SVM} [e) f)].
   Positive fitted values [a) c) e)],
     Negative  fitted values [b) d) f)].
 }
\end{figure}

\section{Conclusions}\label{sec::conclusions}
We have shown how to estimate sentiment  categories proportions for retrieval  through a non aggregative approach, and validating the approach with a  very large result sets.     The non aggregative approach is very efficient and suitable for real time analytics. The method consists in  taking an additive  measure $\mu_\category$ derived from  the classifier  and applied to the entire result set of a query in one single shot. We have also given the cumulative conditions  under which such a measure  can be defined for a given classifier. The model   ignores the category priors, but to compensate this, it learns how to resize the cumulative measure through a linear regression model $\Phi$. We have used  item-driven  and  query-driven settings for $\Phi$, and used three classifiers, two Multinomial Naive Bayes and SVM. We have used the Kolmogorov-Smirnov test to validate the hypothesis that  observed and fitted values come from the same distribution for each method and classifier. In addition, we have used Pearson's correlation test to show that a linearity between the counting and the measures $\mu_\category$ is very strong. The results are also compared  with the  ACC baseline. Both item-driven and query driven approaches work similarly, but only SVM and MNB pass the Kolmogorov-Smirnov test for both positive and negative categories. There is the negative exception of DBM with the item-driven approach. The ACC confirms to  be not stable or reliable both with Pearson correlation and Kolmogorov-Smirnov test.

\small
\bibliography{../../BIBLIOGRAFIE/amati,../../BIBLIOGRAFIE/sentiment,../../BIBLIOGRAFIE/gianni} 
\bibliographystyle{plain}

\end {document}